\catcode`\@=11					



\font\fiverm=cmr5				
\font\fivemi=cmmi5				
\font\fivesy=cmsy5				
\font\fivebf=cmbx5				

\skewchar\fivemi='177
\skewchar\fivesy='60


\font\sixrm=cmr6				
\font\sixi=cmmi6				
\font\sixsy=cmsy6				
\font\sixbf=cmbx6				

\skewchar\sixi='177
\skewchar\sixsy='60


\font\sevenrm=cmr7				
\font\seveni=cmmi7				
\font\sevensy=cmsy7				
\font\sevenit=cmti7				
\font\sevenbf=cmbx7				

\skewchar\seveni='177
\skewchar\sevensy='60


\font\eightrm=cmr8				
\font\eighti=cmmi8				
\font\eightsy=cmsy8				
\font\eightit=cmti8				
\font\eightbf=cmbx8				

\skewchar\eighti='177
\skewchar\eightsy='60


\font\ninei=cmmi9
\font\ninesy=cmsy9

\skewchar\ninei='177
\skewchar\ninesy='60


\font\tenrm=cmr10				
\font\teni=cmmi10				
\font\tensy=cmsy10				
\font\tenex=cmex10				
\font\tenit=cmti10				
\font\tensl=cmsl10				
\font\tenbf=cmbx10				
\font\tentt=cmtt10				
\font\tenss=cmss10				
\font\tensc=cmcsc10				
\font\tenbi=cmmib10				

\skewchar\teni='177
\skewchar\tenbi='177
\skewchar\tensy='60

\def\tenpoint{\ifmmode\err@badsizechange\else
	\textfont0=\tenrm \scriptfont0=\sevenrm \scriptscriptfont0=\fiverm
	\textfont1=\teni  \scriptfont1=\seveni  \scriptscriptfont1=\fivemi
	\textfont2=\tensy \scriptfont2=\sevensy \scriptscriptfont2=\fivesy
	\textfont3=\tenex \scriptfont3=\tenex   \scriptscriptfont3=\tenex
	\textfont4=\tenit \scriptfont4=\sevenit \scriptscriptfont4=\sevenit
	\textfont5=\tensl
	\textfont6=\tenbf \scriptfont6=\sevenbf \scriptscriptfont6=\fivebf
	\textfont7=\tentt
	\textfont8=\tenbi \scriptfont8=\seveni  \scriptscriptfont8=\fivemi
	\def\rm{\tenrm\fam=0 }%
	\def\it{\tenit\fam=4 }%
	\def\sl{\tensl\fam=5 }%
	\def\bf{\tenbf\fam=6 }%
	\def\tt{\tentt\fam=7 }%
	\def\ss{\tenss}%
	\def\sc{\tensc}%
	\def\bmit{\fam=8 }%
	\rm\setparameters\setbaselines\fi}


\font\twelverm=cmr12				
\font\twelvei=cmmi12				
\font\twelvesy=cmsy10	scaled\magstep1		
\font\twelveex=cmex10	scaled\magstep1		
\font\twelveit=cmti12				
\font\twelvesl=cmsl12				
\font\twelvebf=cmbx12				
\font\twelvett=cmtt12				
\font\twelvess=cmss12				
\font\twelvesc=cmcsc10	scaled\magstep1		
\font\twelvebi=cmmib10	scaled\magstep1		

\skewchar\twelvei='177
\skewchar\twelvebi='177
\skewchar\twelvesy='60

\def\twelvepoint{\ifmmode\err@badsizechange\else
	\textfont0=\twelverm \scriptfont0=\eightrm \scriptscriptfont0=\sixrm
	\textfont1=\twelvei  \scriptfont1=\eighti  \scriptscriptfont1=\sixi
	\textfont2=\twelvesy \scriptfont2=\eightsy \scriptscriptfont2=\sixsy
	\textfont3=\twelveex \scriptfont3=\tenex   \scriptscriptfont3=\tenex
	\textfont4=\twelveit \scriptfont4=\eightit \scriptscriptfont4=\sevenit
	\textfont5=\twelvesl
	\textfont6=\twelvebf \scriptfont6=\eightbf \scriptscriptfont6=\sixbf
	\textfont7=\twelvett
	\textfont8=\twelvebi \scriptfont8=\eighti  \scriptscriptfont8=\sixi
	\def\rm{\twelverm\fam=0 }%
	\def\it{\twelveit\fam=4 }%
	\def\sl{\twelvesl\fam=5 }%
	\def\bf{\twelvebf\fam=6 }%
	\def\tt{\twelvett\fam=7 }%
	\def\ss{\twelvess}%
	\def\sc{\twelvesc}%
	\def\bmit{\fam=8 }%
	\rm\setparameters\setbaselines\fi}


\font\fourteenrm=cmr10	scaled\magstep2		
\font\fourteeni=cmmi10	scaled\magstep2		
\font\fourteensy=cmsy10	scaled\magstep2		
\font\fourteenex=cmex10	scaled\magstep2		
\font\fourteenit=cmti10	scaled\magstep2		
\font\fourteensl=cmsl10	scaled\magstep2		
\font\fourteenbf=cmbx10	scaled\magstep2		
\font\fourteentt=cmtt10	scaled\magstep2		
\font\fourteenss=cmss10	scaled\magstep2		
\font\fourteensc=cmcsc10 scaled\magstep2	
\font\fourteenbi=cmmib10 scaled\magstep2	

\skewchar\fourteeni='177
\skewchar\fourteenbi='177
\skewchar\fourteensy='60

\def\fourteenpoint{\ifmmode\err@badsizechange\else
	\textfont0=\fourteenrm \scriptfont0=\tenrm \scriptscriptfont0=\sevenrm
	\textfont1=\fourteeni  \scriptfont1=\teni  \scriptscriptfont1=\seveni
	\textfont2=\fourteensy \scriptfont2=\tensy \scriptscriptfont2=\sevensy
	\textfont3=\fourteenex \scriptfont3=\tenex \scriptscriptfont3=\tenex
	\textfont4=\fourteenit \scriptfont4=\tenit \scriptscriptfont4=\sevenit
	\textfont5=\fourteensl
	\textfont6=\fourteenbf \scriptfont6=\tenbf \scriptscriptfont6=\sevenbf
	\textfont7=\fourteentt
	\textfont8=\fourteenbi \scriptfont8=\tenbi \scriptscriptfont8=\seveni
	\def\rm{\fourteenrm\fam=0 }%
	\def\it{\fourteenit\fam=4 }%
	\def\sl{\fourteensl\fam=5 }%
	\def\bf{\fourteenbf\fam=6 }%
	\def\tt{\fourteentt\fam=7}%
	\def\ss{\fourteenss}%
	\def\sc{\fourteensc}%
	\def\bmit{\fam=8 }%
	\rm\setparameters\setbaselines\fi}


\font\seventeenrm=cmr10 scaled\magstep3		


\newdimen\rp@
\newcount\@basestretchnum
\newskip\@baseskip
\newskip\headskip
\newskip\footskip


\def\setparameters{\rp@=.1em
	\headskip=24\rp@
	\footskip=\headskip
	\delimitershortfall=5\rp@
	\nulldelimiterspace=1.2\rp@
	\scriptspace=0.5\rp@
	\abovedisplayskip=10\rp@ plus3\rp@ minus5\rp@
	\belowdisplayskip=10\rp@ plus3\rp@ minus5\rp@
	\abovedisplayshortskip=5\rp@ plus2\rp@ minus4\rp@
	\belowdisplayshortskip=10\rp@ plus3\rp@ minus5\rp@
	\normallineskip=\rp@
	\lineskip=\normallineskip
	\normallineskiplimit=0pt
	\lineskiplimit=\normallineskiplimit
	\jot=3\rp@
	\setbox0=\hbox{\the\textfont3 B}\p@renwd=\wd0
	\skip\footins=12\rp@ plus3\rp@ minus3\rp@
	\skip\topins=0pt plus0pt minus0pt}


\def\setbaselines{\maxdepth=4\rp@\baselinestretch=\@basestretchnum}


\def\baselinestretch{\afterassignment\@basestretch\@basestretchnum}
\def\@basestretch{%
	\@baseskip=12\rp@ \divide\@baseskip by1000
	\normalbaselineskip=\@basestretchnum\@baseskip
	\baselineskip=\normalbaselineskip
	\bigskipamount=\the\baselineskip
		plus.25\baselineskip minus.25\baselineskip
	\medskipamount=.5\baselineskip
		plus.125\baselineskip minus.125\baselineskip
	\smallskipamount=.25\baselineskip
		plus.0625\baselineskip minus.0625\baselineskip
	\setbox\strutbox=\hbox{\vrule height.708\baselineskip
		depth.292\baselineskip width0pt }}



\def\makeheadline{\vbox to0pt{\baselinestretch=1000
	\vskip-\headskip \vskip1.5pt
	\line{\vbox to\ht\strutbox{}\the\headline}\vss}\nointerlineskip}

\def\makefootline{\baselineskip=\footskip\line{\the\footline}}

\def\big#1{{\hbox{$\left#1\vbox to8.5\rp@ {}\right.\n@space$}}}
\def\Big#1{{\hbox{$\left#1\vbox to11.5\rp@ {}\right.\n@space$}}}
\def\bigg#1{{\hbox{$\left#1\vbox to14.5\rp@ {}\right.\n@space$}}}
\def\Bigg#1{{\hbox{$\left#1\vbox to17.5\rp@ {}\right.\n@space$}}}


\mathchardef\alpha="710B
\mathchardef\beta="710C
\mathchardef\gamma="710D
\mathchardef\delta="710E
\mathchardef\epsilon="710F
\mathchardef\zeta="7110
\mathchardef\eta="7111
\mathchardef\theta="7112
\mathchardef\iota="7113
\mathchardef\kappa="7114
\mathchardef\lambda="7115
\mathchardef\mu="7116
\mathchardef\nu="7117
\mathchardef\xi="7118
\mathchardef\pi="7119
\mathchardef\rho="711A
\mathchardef\sigma="711B
\mathchardef\tau="711C
\mathchardef\upsilon="711D
\mathchardef\phi="711E
\mathchardef\chi="711F
\mathchardef\psi="7120
\mathchardef\omega="7121
\mathchardef\varepsilon="7122
\mathchardef\vartheta="7123
\mathchardef\varpi="7124
\mathchardef\varrho="7125
\mathchardef\varsigma="7126
\mathchardef\varphi="7127
\mathchardef\imath="717B
\mathchardef\jmath="717C
\mathchardef\ell="7160
\mathchardef\wp="717D
\mathchardef\partial="7140
\mathchardef\flat="715B
\mathchardef\natural="715C
\mathchardef\sharp="715D


\def\err@badsizechange{%
	\immediate\write16{--> Size change not allowed in math mode, ignored}}

\baselinestretch=1000
\tenpoint

\catcode`\@=12					
\catcode`\@=11
\expandafter\ifx\csname @iasmacros\endcsname\relax
	\global\let\@iasmacros=\par
\else	\endinput
\fi
\catcode`\@=12


\def\rmb{\seventeenrm}


\def\singlespace{\baselineskip=\normalbaselineskip}
\def\halfspace{\baselineskip=1.5\normalbaselineskip}
\def\doublespace{\baselineskip=2\normalbaselineskip}


\def\AB{\bigskip\parindent=40pt
        \centerline{\bf ABSTRACT}\medskip\halfspace\narrower}
\def\AE{\bigskip\nonarrower\doublespace}
\def\nonarrower{\advance\leftskip by-\parindent
	\advance\rightskip by-\parindent}


\def\boxit#1{\vbox{\hrule\hbox{\vrule\kern3pt
	\vbox{\kern3pt#1\kern3pt}\kern3pt\vrule}\hrule}}

\def\hence{\leavevmode\hbox{\bf .\raise5.5pt\hbox{.}.} }

\def\dalemb#1#2{{\vbox{\hrule height.#2pt
	\hbox{\vrule width.#2pt height#1pt \kern#1pt \vrule width.#2pt}
	\hrule height.#2pt}}}
\def\gtorder{\mathrel{\raise.3ex\hbox{$>$}\mkern-14mu
             \lower0.6ex\hbox{$\sim$}}}
\def\ltorder{\mathrel{\raise.3ex\hbox{$<$}\mkern-14mu
             \lower0.6ex\hbox{$\sim$}}}

\newdimen\fullhsize
\newbox\leftcolumn
\def\twoup{\hoffset=-.5in \voffset=-.25in
  \hsize=4.75in \fullhsize=10in \vsize=6.9in
  \def\fullline{\hbox to\fullhsize}
  \let\lr=L
  \output={\if L\lr
        \global\setbox\leftcolumn=\columnbox\global\let\lr=R \advancepageno
      \else \doubleformat \global\let\lr=L\fi
    \ifnum\outputpenalty>-20000 \else\dosupereject\fi}
  \def\doubleformat{\shipout\vbox{
    \fullline{\box\leftcolumn\hfil\columnbox}\advancepageno}}
  \def\columnbox{\leftline{\vbox{\makeheadline\pagebody\makefootline}}}
  \tolerance=1000 }

\twelvepoint
\doublespace
{\nopagenumbers{
\rightline{IASSNS-HEP-96/104}
\rightline{~~~REVISED April, 1997}
\bigskip\bigskip
\centerline{\rmb Frustrated SU(4) as the Preonic Precursor}
\centerline{\rmb of the Standard Model}
\medskip
\centerline{\it Stephen L. Adler
}
\centerline{\bf Institute for Advanced Study}
\centerline{\bf Princeton, NJ 08540}
\medskip
\bigskip\bigskip
\leftline{\it Send correspondence to:}
\medskip
{\singlespace\leftline{Stephen L. Adler}
\leftline{Institute for Advanced Study}
\leftline{Olden Lane, Princeton, NJ 08540}
\leftline{Phone 609-734-8051; FAX 609-924-8399; email adler@sns.ias.edu}}
\bigskip\bigskip
}}
\vfill\eject
\pageno=2
\AB
We give a model for composite quarks and leptons based on the semisimple 
gauge group $SU(4)$, with the preons in the 10 representation; this choice  
of gauge gluon and preon multiplets is motivated by the possibility of
embedding them in an $N=6$ supergravity multiplet.  
Hypercolor singlets are forbidden in the fermionic sector of this theory; 
we propose that $SU(4)$ symmetry spontaneously breaks to 
$SU(3) \times U(1)$, with the binding of triality nonzero preons and gluons  
into composites, and with the formation of a color 
singlet condensate that breaks the initial $Z_{12}$ vacuum symmetry to 
$Z_{6}$.  The spin 1/2 fermionic composites have the triality structure 
of a quark lepton family, and the initial $Z_{12}$ symmetry implies that 
there are six massless families, which mix to give three distinct families, 
two massless with massive partners and one with both states massive, 
at the scale of the condensate.  The spin 1 triality zero composites of the 
color triplet $SU(4)$ gluons, 
when coupled to the condensate and with the color singlet representation of 
the 10 acting as a doorway state, lead to weak interactions of the fermionic 
composites through an exact 
$SU(2)$ gauge algebra.  The initial $Z_{12}$ symmetry implies that this 
$SU(2)$ gauge algebra structure is doubled, which in turn requires that the 
corresponding independent gauge bosons must couple to chiral components 
of the composite fermions.  Since the 
$U(1)$  couples to the 10 representation as $B-L$, 
an effective $SU(2)_L \times SU(2)_R \times
U(1)_{B-L}$ electroweak theory arises at the condensate scale, 
with all composites having the 
correct electric charge structure.  A renormalization group analysis shows 
that the conversion by binding of one 10 of $SU(4)$ to 12 triplets of $SU(3)$  
 
can give a very large, calculable hierarchy ratio between the $SU(4)$  and  
the hadronic mass scales.  

\AE
\bigskip\bigskip
Return-Path: adler@sns.ias.edu
Received: from thunder.sns.ias.edu (thunder.sns.ias.edu [198.138.243.12]) by 
lonestar.sns.ias.edu (8.6.12/8.6.12) with ESMTP id PAA10717 for 
<val@sns.ias.edu>; Thu, 1 May 1997 15:34:30 -0400
Received: from thunder (adler@localhost) by thunder.sns.ias.edu (8.7.4/8.6.12) 
with ESMTP id PAA15321 for <val>; Thu, 1 May 1997 15:34:22 -0400 (EDT)
Message-Id: <199705011934.PAA15321@thunder.sns.ias.edu>
X-Authentication-Warning: thunder.sns.ias.edu: adler owned process doing -bs
X-Mailer: exmh version 1.6.7 5/3/96
To: val@sns.ias.edu
Subject: revised 9610190 - please post to bulletin board (originated from you)
Mime-Version: 1.0
Content-Type: text/plain; charset=us-ascii
Date: Thu, 01 May 1997 15:34:20 -0400
From: Stephen Adler <adler@sns.ias.edu>

\vfill\eject
\pageno=3
\centerline{{\bf 1.~~Introduction}}

Although the repetition of quark-lepton families is strongly 
suggestive of composite structure,  no plausible composite model has yet 
emerged, and the current focus of research on unification is based on 
the alternate idea of grand unification.  In this paper we reexamine 
the idea of compositeness in the context of a new model for the 
formation of composites.  Nearly all work to date on composites 
has assumed a ``QCD-like'' paradigm, in which the preons couple to a 
hypercolor force field, that acts independently of the standard model gauge 
fields (by a group theoretic direct product) and binds the 
preons into hypercolor singlets.  Although based on well-studied physics, 
this approach suffers from a serious problem relating to chiral symmetry.  
In general, the direct product structure leads to a large global chiral 
symmetry group, even after the breaking of the overall $U(1)$ chiral 
symmetry by instantons.  As a consequence, the 't Hooft anomaly matching 
conditions [1] must be obeyed if massless composites are to be possible, and 
extensive searches for solutions to these equations [2] show none that match  
the observed particle spectrum.  When the 't Hooft conditions are not 
obeyed, a chiral symmetry breaking condensate must form on the 
binding scale of the theory, as happens in QCD; the composites then get   
large masses, and composite structure with a large hierarchy ratio is 
not possible.   

We propose in this paper an alternative approach to composite structure, 
based on a partial grand unification using a hypercolor group and preonic 
fermion multiplet structure for which the hypercolor forces are 
{\it frustrated} in the fermionic sector, in the sense that hypercolor 
singlet fermionic states are forbidden.   
Together with a proposed chiral symmetry breaking chain, this leads to 
quark lepton composites (and further matter bound states of these), in which 
the impossibility of fermionic hypercolor singlets translates into 
the participation of all fermionic composite states in spin-1 gluon  
mediated gauge interactions, as experimentally observed.  

Although we do not make explicit use of  
supersymmetry in this paper, supersymmetry ideas are a principal motivation  
in the formulation of our model.  Specifically, the hypercolor 
gauge group $SU(4)$, with the preons in a single 10 representation of 
Dirac fermions, are chosen for study because they correspond to the only  
$SU(N)$ based generalized ``rishon'' model that is embedable in an 
extended supergravity multiplet.  Similarly, the chiral symmetry breaking 
chain that we postulate is strongly motivated by recent results of 
Seiberg [3] (for reviews see [4]) showing that there are supersymmetric 
systems in which preons with vector-like gauge couplings 
can form massless fermion composites, thus contradicting the 
``most attractive channel'' rule for chiral symmetry breaking, and also 
showing that in supersymmetric systems massless composite non-Abelian gauge 
gluons can occur.  With these recent results in mind, we propose 
a symmetry breaking route by which the known structure of the standard model 
emerges emerges from our preon model, with the correct family multiplicity, 
but (as distinguished from conventional grand unification) with the 
intermediate boson states arising as composites of three fundamental gluons.  
It is this latter feature that allows us to evade the usual  
grand unification restriction requiring a unification group of at 
least rank 4 (since the standard model is rank 4), and permits a hybrid 
grand-composite unification in the rank 3 group $SU(4)$.  This in turn    
makes possible renewed consideration of the appealing idea that all 
matter and force carriers, including gravitation, may lie in a 
single extended supergravity multiplet.

\bigskip
\centerline{{\bf 2.~~Counting relations for generalized ``rishon'' models~~}}

Since our construction uses some of the basic notions of the 
Harari-Seiberg [5] 
``rishon'' model version of the Harari-Shupe [6] scheme, we briefly explain 
the 
relevant aspects here.  The original Harari-Shupe scheme hoped to generate 
$SU(3)$ symmetry as a permutation symmetry acting on the preons, which is not 
possible within standard complex quantum field theory; moreover, despite an 
extensive investigation [7] that we have carried out 
of noncommutative quaternionic quantum 
mechanics, we still have found no concrete way to realize a dynamically 
generated 
exact  color symmetry.  So instead, 
following Harari and Seiberg, we will assume that the $SU(3)$ color 
group is present as a subgroup in the fundamental gauge interactions, which 
are treated in standard quantum field theory.  

In the Harari-Seiberg model, the fundamental preons are postulated to be 
``rishon'' states $T$ and $V$, and their antiparticle states $\overline{T}$ 
and $\overline{V}$, with electric charges $Q$ and $SU(3)$ trialities 
$Tri$ assigned as follows:  
$$\eqalign{
Q(T)=&\,1/3,~~Tri(T)=1,~~~~Q(\overline{T})=-1/3,~~Tri(\overline{T})=-1\cr 
Q(V)=&\,0,~~~~Tri(V)=-1,~~~Q(\overline{V})=0,~~~~~~Tri(\overline{V})=1.\cr  
}\eqno(1)$$
The quark and lepton states in the first family are then constructed as 
three preon composites according to the scheme 
$$\eqalign{
e^+=&\,TTT \cr
u=&\,TTV\cr
\overline{d}=&\,TVV\cr
\nu=&\,VVV~,~~~\cr
}\eqno(2)$$
with the corresponding expressions for $e^-,\overline{u},d,\overline{\nu}$
obtained by replacing $T,V$ by $\overline{T}, \overline{V}$.  Defining 
the particle number $n_T,~n_V,~n_e,~n_u,~n_d,~n_{\nu}$ as the difference 
between the number of particles and antiparticles of the indicated type 
(counting $e^-$, as usual, as a particle) we immediately find from Eqs.~(1) 
and (2) 
the following counting relations
$$\eqalign{
n_T=&-3n_e+2n_u-n_d\cr
n_V=&~~\,~3n_{\nu}-2n_d+n_u ~~~,\cr
}\eqno(3a)$$
from which we find
$$\eqalign{
{1\over 3}(n_T-n_V)=&{1\over 3}(n_u+n_d)-(n_e+n_{\nu})  \cr
{1\over 3}(n_T+n_V)=&\,n_u-n_d+n_{\nu}-n_e~.\cr
}\eqno(3b)$$
Since Eq.~(1) implies that the electric charge $Q$ is given by 
$$Q={1\over 3}n_T={1\over 6}(n_T+n_V)+{1\over 6}(n_T-n_V)~,~~~\eqno(4a)$$
if we define the baryon number $B$, lepton number $L$, and 
preon or fermion number $F$  by 
$$\eqalign{
B=&{1\over 3}(n_u+n_d)\cr
L=&\,n_e+n_{\nu}\cr
F=&\,n_T+n_V~,~~~\cr
}\eqno(4b)$$
the electric charge can be rewritten as 
$$Q= {1\over 6}F +{1 \over 6}(n_T-n_V)=
{1\over 6}F +{1 \over 2}(B-L)~.\eqno(4c)$$
Finally, making the definitions 
$$\eqalign{
I_{3L}=&{1 \over 2}[n_{uL}-n_{dL}+n_{\nu L}-n_{eL}]\cr
I_{3R}=&{1 \over 2}[n_{uR}-n_{dR}+n_{\nu R}-n_{eR}]~,~~~\cr
}\eqno(5a)$$
where $R,L$ denote left, right helicity components, we can rewrite 
Eq.~(4c) by virtue of Eqs.~(4b) and (3b) as
$$Q=I_{3L}+I_{3R}+{1 \over 2}(B-L)~.\eqno(5b)$$

>From these manipulations, and the requirement that the electric charge  
$Q$ be conserved, we can draw two general conclusions about any 
preonic scheme obeying the charge, triality, and composite state 
assignments of Eqs.~(1) and (2).  From the first equality in Eq.~(4c),  
we learn that at the preonic level, the electric charge must be constructed 
from the ungauged conserved total preon number $F$ and a conserved 
gauged $U(1)$ charge 
proportional to $n_T-n_V$.   From Eq.~(5b), we learn that in an effective 
gauge theory of the composites, the electric charge necessarily has the 
form found in left-right symmetric $SU(2)_L \times SU(2)_R \times 
U(1)_{B-L}$ electroweak models [8], which by a well-understood symmetry 
breaking mechanism can give the standard 
$SU(2)_L \times U(1)_Y$ electroweak model, and so we expect to make the 
connection with standard model physics by this route.  
\bigskip
\centerline{\bf 3.~~The role of the $10$ and $15$ representations of $SU(4)$}
                                              
Since the generalized rishon model described in the previous section is
a calculus of $SU(3)$ trialities, any embedding of $SU(3)$ in a larger 
gauge group can potentially give a preon model of this type.  An extensive 
tabulation of color $SU(3)$ embeddings has been given in the review of 
Slansky [9], and there is clearly a plethora of possible models.  To 
narrow the field, we introduce a ground rule derived from supersymmetry 
considerations:  We will only consider gauge groups that are 
potentially embedable in an extended Poincar\'e supermultiplet.  
An examination of the spin 1 content of such multiplets [10] shows 
that the only possibilities admitting an $SU(3)$ embedding 
are the $SO(N)$ gauge groups with $N=6,7,8$;  
we will choose as our candidate model the smallest of these,
$SO(6) \sim SU(4)$, which has an adjoint multiplet of $15$ gauge gluons and   
has long been considered [11] a possible unification group.
For $N=6$, extended supergravity requires that the fermions come 
in multiplets of either 6 or 20 two-component states; 
an examination of possible $SU(4)$ representations corresponding to 
these numbers shows that the only case leading to a satisfactory 
representation of the triality and charge rules of Eq.~(1) 
corresponds to putting  
the fermions in left handed $10$ and $\overline{10}$ 
representations of $SU(4)$.   
>From the viewpoint of a possible supergravity extension it is very 
encouraging that, as discussed below in Sec. 4, these representations 
correspond {\it precisely} to the decomposition of the antisymmetric
tensor representations 15 and 20 of $SU(6)$, the automorphism group used 
[12] in constructing the $N=6$ extended Poincar\'e multiplet, 
under the irregular embedding of the $SU(4)$ subgroup. 

The specific properties of $SU(4)$ that are needed for our model are given in 
Tables 25-27 of Slansky [9], and some related properties of
$SU(3)$ that we use are given in Tables 23 and 24.  As motivated in the 
preceding paragraph, our model consists of 
a $10_L$ and a $\overline{10}_L$ of two-component Weyl spinor preons (or 
equivalently, of a single 10 of Dirac four-component 
fermionic preons) gauged by 
the 15 adjoint representation of $SU(4)$.  
Since our model is vectorlike, there are no gauge anomalies and 
the model is renormalizable.  Because the 10 representation has quadrality 2, 
the $\overline{10}$ has quadrality $-2\equiv 2$ modulo 4, and so 
any odd number of preons has quadrality 2 modulo 4. Thus fermionic 
composites can never be $SU(4)$ singlets, and so the hypercolor forces in   
the fermionic sector of 
the model are frustrated [13].  

We postulate that as a consequence, the $SU(4)$   
symmetry is spontaneously broken into the maximal $SU(3) \times U(1)$ 
subgroup by a condensate, to be discussed further in Sec. 4, characterized 
by an energy scale {\it that is much smaller} than the characteristic 
scale $\Lambda_H$ at which the $SU_4$ running coupling becomes strong.  
Even at the scale $\Lambda_H$, a very small asymmetric perturbation is 
enough to specify the favored subgroup into which the $SU(4)$ symmetry 
breaks, and moreover, when the running coupling becomes large, small 
asymmetries can be amplified and have a decisive effect on the dynamics.  
Thus our operating assumption will be that 
at the scale $\Lambda_H$ where the $SU(4)$ dynamics becomes strongly 
coupled, the theory reorganizes itself according to the $SU(3)$ content 
of the $SU(4)$ multiplets.  More specifically, we shall assume that {\it in 
each sector characterized by definite preon number and $SU(3)$ triality 
the fundamental fields with  triality nonzero bind 
to form composites characterized by the smallest $SU(3)$  Casimir available 
for that triality.}   Let us now explore the consequences of this   
assumption.  

We begin with the gluon 15 multiplet, which under $SU(4) \supset SU(3) \times 
U(1)$  decomposes as
$$15=1(0)+3(-1/3)+\overline{3}(1/3)+8(0)~~~,\eqno(6)$$
where the numbers in parentheses are the $U(1)$ charges, for which we 
adopt a normalization that is $1/4$ of that used in Slansky's 
Table 27.  We see that the multiplet contains 9 gluons that are 
$U(1)$ neutral: the $U(1)$ force carrier  1(0) and the 8 $SU(3)$ force 
carriers  8(0).  If we imagine all $SU(4)$ states displayed on a three
dimensional plot with the $U(1)$ charge running along the $z$ axis, 
then these 9 gluons lie in the $xy$ plane, and we shall refer to them 
collectively as ``horizontal gluons''.  The remaining 6 gluons of $SU(4)$ 
consist of two $SU(3)$ triplets $3$ and $\overline{3}$, with $U(1)$ 
charges $\pm 1/3$; we shall refer to these collectively as the 
``vertical gluons''.\footnote{*}{The familiar plot of the $SU(4)$ 
pseudoscalar meson 16-plet [14] takes 
this form, when the $\pi,\eta,K$ mesons are relabeled 
``horizontal gluons'' and the $D$ mesons are relabeled ``vertical gluons''.}  
According to our fundamental assumption, in the 
sector with $U(1)$ charge 1 and angular momentum 1, 
three vertical gluons in the $\overline{3}$ representation will bind, 
picking up horizontal gluon components of the wave function as needed, 
to form an $SU(3)$ singlet composite vector meson state with $U(1)$ 
charge 1; similarly, the three $3$ gluons will bind to form 
a vector state with $U(1)$ charge $-1$.  These composite vector states will  
ultimately play the role of components of the 
intermediate vector bosons $W^{\pm}$.  Thus, our picture is that 
the adjoint representation of the $SU(4)$ group ``folds'' in a manner 
dictated by the decomposition into $SU(3) \times U(1)$, with horizontal 
gluons remaining as massless gauge gluons, but with the vertical gluon 
triplets binding to form two color $SU(3)$ singlets that will become the 
charged weak force  carriers.

We turn next to the $10_L$ and $\overline{10}_L$ representations, 
which under $SU(4) \supset SU(3) \times U(1)$  decompose as
$$\eqalign{
10_L=&\,1_L(1/2)+3_L(1/6)+6_L(-1/6)\cr
\overline{10}_L=&\,1_L(-1/2)+\overline{3}_L(-1/6)+\overline{6}_L(1/6)~~~,\cr
}\eqno(7)$$
where the $U(1)$ charge normalization is again $1/4$ of that used by 
Slansky.  Since the $SU(3)$ 6 representation has triality $2\equiv -1$ 
modulo 3, if we assign states $S_L$, $T_L$, and $V_L$  to the $\overline{10}$ 
of $SU(4)$, and their antiparticles correspondingly to the 10, according to 
$$\eqalign{
S_L\equiv &\, 1_L(-1/2)~,~~~T_L\equiv\overline{6}_L(1/6)~,~~~ 
V_L\equiv\overline{3}_L(-1/6)  \cr
\overline{S}_L\equiv &\, 1_L(1/2)~,~~~\overline{T}_L\equiv 6_L(-1/6)~,~~~ 
\overline{V}_L\equiv 3_L(1/6)~~~,  \cr
}\eqno(8)$$  
then the $T$ and $V$ states have the trialities required by Eq.~(1).  
Furthermore, since the ungauged preon number 
$n_{10}=n_{10_L}-n_{\overline{10}_L}$ and the $U(1)$ 
charge $Q_{U(1)}$ are both conserved, a satisfactory definition of the 
conserved electric charge operator $Q$ at the preonic level is 
$$Q=-{1\over 6}n_{10}+Q_{U(1)}~~~,\eqno(9)$$
according to which the charges of $S$, $T$, and $V$ are
$$
Q_{S}={1\over 6}-{1\over 2}=-{1\over 3}~,~~~Q_{T}={1\over 6}+{1\over 6}
={1\over 3}~,~~~Q_{V}={1\over 6}-{1\over 6}=0~~~,\eqno(10)$$
in agreement (for $T$ and $V$) with the charge assignments of Eq.~(1).  
Therefore the $T$ and $V$ preonic states in the 10 and $\overline{10}$ 
give a realization of the ``rishon'' model.  According to our fundamental 
assumption, in the three preon angular momentum $1/2$ 
sector the $T$'s and $V$'s will bind, picking up horizontal gluons as needed,  

into the states with lowest $SU(3)$ Casimir for each triality (the 
$S$'s are color singlets and so do not directly participate), giving color 
triplet quarks and color singlet leptons according to the scheme of 
Eq.~(2).

In addition to the three preon composites of Eq.~(2), one in 
principle can have fractionally charged triality zero 
composites consisting of two preons 
and one antipreon, or of one preon and two antipreons, such as  $TT
\overline{V}$ and $T\overline{V}\overline{V}$.  These states are discussed 
in Appendix B, where it is argued from the anomaly structure of the theory 
that they cannot simultaneously be present, along with the composites 
of Eq.~(2), in a gauged electroweak low energy effective action.  
\bigskip
\centerline{\bf 4.~~Chiral symmetry structure of the model}

We turn next to the crucial issue of the chiral symmetry structure of the 
model, and its implications for the spectrum of massless particles.   
According to what, until recently, was standard lore about chiral symmetry 
breaking based on perturbative studies supplemented by instanton [15] 
and lattice gauge 
theory arguments [16], a hypercolor gauge theory containing fermions 
with vector-like 
couplings would be expected to follow the ``most attractive channel'' [17] 
rule obeyed in QCD.  According to this rule, one would expect a 
fermion-antifermion condensate (a $Z_2$ condensate in the terminology used 
below) to 
form, giving the composites masses at the hypercolor scale, and thus 
precluding their identification with standard model quarks and leptons.    
However, as noted in Sec. 1, recent results of Seiberg [3] (for expository 
reviews see Seiberg [4] and Intriligator and Seiberg [4]) 
show that by using holomorphy information in 
supersymmetric theories, one can find 
examples that contradict the most attractive channel rule.  
Specifically, these authors study supersymmetric QCD with $N_f$ 
quarks in the fundamental and antiquarks in the anti-fundamental 
representation, which is a model in which all fermions have vectorlike 
couplings.   This theory is asymptotically free for $N_f < 3 N_c$.  For 
$N_f=N_c$ the theory confines and breaks chiral symmetry, as expected 
from the most attractive channel rule, but for $N_f=N_c+1$ (see Sec. 4.3 of 
Intriligator and Seiberg [4]) the theory 
exhibits confinement into composites without chiral symmetry breaking, 
contradicting the conclusion one would get from the most attractive 
channel rule.  We interpret this example as indicating that the older 
standard lore about chiral symmetry breaking must be treated skeptically when 
dealing with supersymmetric  theories.  

Although the model of this                    
paper, in a non-supersymmetric context, might be expected to follow the 
most attractive channel rule, as we have discussed above the model is 
motivated by the possibility of an $N=6$ supersymmetric embedding.  In 
the context of such an embedding, we take the results of [3, 4] 
to indicate that we contradict no known results in quantum 
field theory by postulating that chiral symmetry breaking in our model 
does not follow 
the most attractive channel rule, and that a $Z_2$ condensate does not 
form at the hypercolor scale.  Instead, we assume that the only chiral 
symmetry breaking relevant at the hypercolor scale is that required by the 
instanton induced effective potential, which we now proceed to analyze,  
with further breaking of chiral symmetry occurring only at energies much 
below the hypercolor scale.

Because the fermion representation structure in our model is simply  
$10_L + \overline{10}_L$, there are only two global symmetry currents.  
The first, 
$$V_{\mu}=\overline{\psi}_{10_L} \gamma_{\mu} \psi_{10_L}-  
\overline{\psi}_{\overline{10}_L} \gamma_{\mu} \psi_{\overline{10}_L}
~~~,\eqno(11a)$$
is the conserved vector preon number current, whose conserved charge is 
$n_{10}$.  The second, 
$$X_{\mu}=\overline{\psi}_{10_L} \gamma_{\mu} \psi_{10_L}+
\overline{\psi}_{\overline{10}_L} \gamma_{\mu} \psi_{\overline{10}_L}
~~~,\eqno(11b)$$
is an axial current, the conservation of which is broken by the $U(1)$  
axial anomaly.  The effect of $SU(4)$ instantons, combined with the 
$U(1)$ anomaly, is to induce an effective chiral symmetry breaking potential 
which has the structure 
$$\Delta V\sim {\rm constant} \times 
[\psi_{10_L}\psi_{\overline{10}_L}]^6~~~,\eqno(12)$$
with $SU(4)$ and Lorentz indices contracted to form singlets (the details 
of this do not concern us), and with the exponent in Eq.~(12) taking the 
value 6 because this is the index $\ell(10)$ 
of the 10 representation given in the Slansky [9] tables, which is one half 
the number of fermion zero modes [15] in an instanton background.\footnote{*}
{Equation (4.31) of Peskin [15] writes the exponent of Eq.~(12)  
as $k=n_{10}C(10)+n_{\overline{10}}C(\overline{10})$, 
with Peskin's $n_{10}=n_{\overline{10}}=1$ the number of Weyl spinors, 
and with Peskin's $C(10)=C(\overline{10})=3=\ell(10)/2$ 
one half of the Slansky index $\ell(10)$.}
Under a chiral transformation of the left-handed fields,
$$\eqalign{
\psi_{10_L} \to &\,  \psi_{10_L} \exp(i\alpha)  \cr
\psi_{\overline{10}_L} \to &\, \psi_{\overline{10}_L} \exp(i\alpha) ~~~,\cr
}\eqno(13a)$$
the symmetry breaking potential of Eq.~(12) transforms as
$$\Delta V \to \Delta V \exp(12 i \alpha)~~~, \eqno(13b)$$
and so breaks the continuous chiral symmetry of the Lagrangian down to 
a discrete $Z_{12}$ subgroup
$$\alpha = {2 \pi k \over 12} ~~~,\eqno(13c)$$
with $k$ any integer.  

We now follow a remark of Weinberg [18] that a $Z_K$ 
chiral symmetry can protect certain composites from acquiring 
masses, together with a 
suggestion of Harari and Seiberg [19], that a $Z_K$ chiral 
symmetry can be used as a quantum number to distinguish between different 
quark lepton families. 
Because there are no conserved global axial vector currents in the model, 
there are no 't Hooft anomaly matching conditions to be satisfied, and so 
this $Z_K$ analysis, with $K=12$ in our case, is the sole criterion 
governing the appearance of low mass composite fermion states.  We begin by 
noting that any fermionic composite is constructed from an odd number 
of fields $\psi_{10},~\psi_{\overline{10}}$ or their adjoints; therefore 
(since,  if the sum of two integers is odd, so is their difference) the 
general composite monomial, which we denote by the Dirac spinor $\Psi_N$, 
transforms with  
an odd power $2N+1$ under the $Z_{12}$ chiral transformation of Eqs.~(13a-c), 
$$\Psi_N \to  \exp[i 2 \pi  \gamma_5 (2N+1) k /12]\Psi_N ~~~,\eqno(14a)$$  
which, since $\gamma_5$ has eigenvalues $1,-1$ on $L,R$ chiral states, can 
be written as
$$\Psi_{N\, L,R} \to  \exp[i 2 \pi (1,-1)  (2N+1) k /12]
\Psi_{N\, L,R} ~~~.\eqno(14b)$$  
Let us now analyze the circumstances under which 
(i) two different 
monomials $\Psi_N,~\Psi_M~, N \neq M$ 
behave as members of distinct families, and   
(ii) the monomial  $\Psi_N$ is protected from acquiring a mass.

Assuming that the low energy effective theory remains a gauge theory, 
two different monomials $\Psi_N$ and $\Psi_M$ will represent distinct 
families if the chiral transformation of Eqs.~(14a, b) forbids the occurrence 
of off-diagonal kinetic energy and gauge coupling terms
$$\Psi^{\dagger}_M \gamma_0\gamma_{\mu}\partial^{\mu} \Psi_N ~,~~~
\Psi^{\dagger}_M \gamma_0\gamma_{\mu} \Psi_N A^{\mu}~~~, \eqno(15a)$$ 
with $A^{\mu}$  a chirally invariant gauge potential.    
Thus the occurrence  
of the couplings of Eq.~(15a) will be forbidden if 
$$[2N+1 -(2M+1)] k/12=(N-M)k/6   \eqno(15b)$$ 
takes a noninteger value for any integer $k$.  Clearly, for $N-M=1,2,3,4,5$, 
Eq.~(15b) is not an integer for $k=1$, while when $N-M=6P$ with P an integer, 
Eq.~(15b) is integer-valued for all $k$.  Hence the $Z_{12}$ invariance 
of our theory implies that there are exactly 6 fermion families.   
Let us consider next what happens when the $Z_{12}$ invariance group is  
successively broken by condensates to the subgroups $Z_6$, $Z_4$, and $Z_2$.
For $Z_6$, the 6 in Eq.~(15b) is replaced by 3, and the same reasoning 
implies 
that there are 3 fermion families, labeled by $N=0,1,2$, 
the first family corresponding to a mixture of
$N=0,3$ of the original 6, the second to a mixture of $N=1,4$ and the 
third to a mixture of 
$N=2,5$.  When $Z_{12}$ is broken to $Z_4$, the 6 in Eq.~(15b) is replaced 
by 2, and our reasoning implies that there are 2 fermion families, labeled by 
$N=0,1$, 
corresponding to respective mixtures of the 
even and of the odd $N$ values in the original 6.  Finally, 
when $Z_{12}$ is broken to $Z_2$, the 6 in Eq.~(15b) is replaced by 1, and 
only 1 distinct family remains, in other words, all of the original 6 
values of $N$ are allowed to mix.  

Now that we have determined the family structure, let us analyze when 
the families are required by the discrete chiral symmetry to remain 
massless.  In carrying out this analysis, we will make the technical 
assumption that we only have to consider mass terms within each distinct 
family as characterized by the kinetic energy and gauge term analysis just 
given, and that we can ignore possible mass terms linking families that 
cannot couple through the kinetic and gauge coupling 
Lagrangian terms of Eq.~(15a).  As 
discussed in Appendix C, this assumption is equivalent to the assumption 
that the Coleman-Mandula [20] analysis of symmetries of the $S$ matrix
extends to the case in which the internal symmetry is a discrete chiral 
symmetry as in Eq.~(13), rather than a continuous symmetry.  With this 
assumption, 
the monomial $\Psi_N$ in general can acquire a Dirac mass if  
$$m_D \Psi^{\dagger}_{N\, R} \gamma_0 \Psi_{N\, L} +{\rm adjoint}  
\eqno(16a)$$ 
is invariant under the chiral transformation of Eqs.~(14a, b),   
and (if neutral) it can acquire a Majorana mass [8] if 
$$ m_L \Psi^T_{N\, L} C^{-1}\Psi_{N \, L} 
+m_R \Psi^T_{N\, R} C^{-1}\Psi_{N \, R} +{\rm adjoint} ~~~,\eqno(16b)$$
with $C$ the charge conjugation matrix, is similarly invariant. (In the  
above equations, $m_D,~m_R,~m_L$ denote arbitrary complex constants.)  
Therefore,  $\Psi_N$ remains 
massless only if 
$$2 (2N+1) k/12  =(2N+1) k/6  \eqno(16c)$$
is not an integer for some integer $k$.  Clearly, to analyze this condition 
we need only keep the residue of $N$ modulo 3, so that we have  the three 
cases $N=0,1,2$ to consider.  For $N=0$ we have $2N+1 =1$ and for $N=2$ 
we have $2N+1=5\equiv -1$ modulo 6, and so in both of these cases Eq.~(16c) 
is 
noninteger as long as $k$ is not divisible by 6.  For $N=1$, Eq.~(16c) 
reduces to $k/2$ and is noninteger for odd $k$, and integer for even $k$.
Hence in all cases the composite $\Psi_N$ is required to remain massless 
by the $Z_{12}$ invariance.   

We next consider what happens to the mass 
analysis when $Z_{12}$ is broken into one of its subgroups.  When $Z_{12}$ 
is broken to $Z_6$, the 6 in Eq.~(16c) is replaced by 3.  In this case, 
for $N=0$ we still have $2N+1=1$, and for $N=2$ we have $2N+1=5\equiv -1$ 
modulo 3, and so in both of these cases Eq.~(16c) remains a noninteger 
as long as $k$ is not divisible by 3, and the corresponding $\Psi_N$ is
required to remain massless.  On the other hand, when $N=1$ we have 
$(2N+1)k/3=k$, which is always an integer, so the $N=2$ family is no longer
protected by discrete chiral invariance from acquiring a mass.  
Thus of the three distinct families left when $Z_{12}$ is broken to $Z_6$, 
two remain massless, but the composites in the remaining one 
can acquire masses.  When $Z_{12}$  
is broken to $Z_4$, the 6 in Eq.~(16c) is replaced by 2.  We now have 
$(2N+1)k/2=Nk+k/2$, which is always nonintegral for odd $k$ irrespective of 
the 
value of $N$.  Hence the two distinct families left when $Z_{12}$ is 
broken to $Z_{4}$ remain massless.  Finally, when $Z_{12}$ is broken to 
$Z_2$, the 6 in Eq.~(16c) is replaced by 1, giving $(2N+1)k$ which is always 
an integer.  The composites in the one remaining distinct family can then 
acquire masses.    

To summarize, we have shown that a symmetry breaking chain 
$$Z_{12} \to Z_6 \to Z_2 \eqno(17a)$$   
leads to a family mixing and mass generation chain 
$$0,1,2,3,4,5 \to (03),(25),(14)_M \to (012345)_M ~~~,
\eqno(17b)$$
while a symmetry breaking chain 
$$Z_{12}  \to Z_4 \to Z_2 \eqno(17c)$$   
leads to a family mixing and mass generation chain 
$$0,1,2,3,4,5 \to  (024),(135) \to (012345)_M ~~~,
\eqno(17d)$$
where the numbers correspond to the $N$ values of the original 6 massless 
families enforced by $Z_{12}$, where families with $N$ values in parentheses   

mix, and where a subscript $M$ indicates that there is no discrete chiral 
protection against acquiring a mass.  We shall show in the next 
section that in our model, Lorentz scalar, electrically neutral, color 
singlet condensates exist corresponding to each step in the chains of 
Eqs.~(17a-d).  Thus our model 
is capable of generating the kind of family structure observed in the 
standard model:  At the $Z_6$ level, there are three distinct families, 
two of which are protected from acquiring mass; at the $Z_4$ level two 
massless families remain, but they mix with the third heavy family; 
while at the $Z_2$ level the three $Z_6$ or two $Z_4$ 
families mix to form one family, all 
states of which can acquire mass. 
We shall leave to a future investigation 
a detailed study of the mass and mixing matrices that arise from this 
scheme.  However, even without a detailed analysis, it is clear that 
for the model to be phenomenologically viable,  the scale corresponding 
to the breaking $Z_{12} \to Z_6$ must be in the electroweak range 
200 GeV to 1 TeV, and {\it our model then makes the prediction 
that there should 
be heavy duplicates of each fermion family in this range}.  
These are not expected to be mirror fermions, but 
rather should have weak couplings of the same type as their currently 
observed partners.  Once we have associated the $Z_6$ condensate with the 
electroweak scale, our model implies on general grounds that 
breaking to  $Z_2$  
{\it  must} occur at the QCD scale, because the 't Hooft anomaly matching 
conditions for quark binding into color singlets are not obeyed when more  
than one family is initially present [21]. (Of course, detailed QCD studies  
indicate that a chiral symmetry breaking $Z_2$ condensate appears even when 
only one family is present, although not required [21] by the anomaly 
matching conditions, which are satisfied for one family.)

To conclude this section, let us address the wider question of what groups 
and fermion representations could permit an analogous chain leading to 3 
families.  
Clearly, for a representation with index $\ell$, the analog of Eq.~(15b) 
implies that the number of initially massless 
distinct families is $\ell$.  Our analysis shows that it is not possible to 
get 3 massless families without heavy partners, since when $\ell=3$ 
we found that only 2 of the 3 families were protected from acquiring masses.  
In fact, among all the Lie groups catalogued by Slansky [9], only the group 
$SU(5)$ has a representation, the 10, with index 3.  Under 
$SU(5) \supset SU(3) \times SU(2) \times U(1)$, the 10 decomposes as 
$$10=(1,1)(6) +(3,2)(1)  +(\overline{3},1)(-4)  ~~~,\eqno(18a)$$
and so if we define the charge as 
$$Q={1 \over 15}[-n_{10}+Q_{U(1)}]~~~,\eqno(18b)$$
we get two representations with the charge and helicity assignments 
assumed in Sec. 2.  However, there are problems -- since five 10's 
can bind to form a hypercolor singlet, there is no reason for non-singlet  
composites to dominate, and also, the $V$ is an $SU(2)$ doublet while the    
$T$ is an $SU(2)$ singlet.  Nonetheless, this model deserves further study.  

Turning to groups catalogued by Slansky 
with representations with index 6, there are the 
8 of $SU(3)$, the 10 of $SU(4)$, the 20 of $SU(6)$, and the 28 of $SU(8)$.  
The first of these 
does not have the triality structure needed in Sec.~2; the second is the 
model of this paper; and the third is related to the model of this paper 
as follows.  The sextality of the 20 of $SU(6)$ is 3, and so in a model        
 
based on this representation, the $SU(6)$ forces 
are also frustrated in the fermionic sector.  
Under the irregular embedding $SU(6) \supset SU(4)$, 
the 20 of $SU(6)$ decomposes as [22]
$$20[SU(6)]=10[SU(4)] + \overline{10} [SU(4)]~~~,
\eqno(19)$$
and the 15 of $SU(6)$ decomposes as 
$$15[SU(6)]=15[SU(4)]~~~.\eqno(20)$$
This indicates that 
our model can also be obtained from $SU(6)$, a fact that is relevant 
for possible supergravity embeddings.  Finally, we consider the 
28 of $SU(8)$.  Under $SU(8) \supset SU(3) \times SU(5) \times U(1)$, this 
representation decomposes as $28=(1,10)(6) +(3,5)(-2) +(\bar 3,1)(-10)$, 
so we 
again encounter the problem that the $3$ and the $\bar 3$ of $SU(3)$ 
transform 
under different representations of the other non-Abelian factor group 
[$SU(5)$ 
in this case] in the decomposition.  

One could also consider groups with representations with index $\ell$ 
larger than 6 but still divisible by 3.  These include the 27 of $SO(7)$ 
with index 18, the 28 of $SO(8)$ with index 12, the 54 of $SO(10)$ with 
index 24, the 52 of $F_4$ with index 18, the 78 of $E_6$ with index 24, 
the 133 of $E_7$ with index 36, and the 248 of $E_8$ with index 60.  
However, for $\ell=3k$ massless families with $k>2$, the number of fermions 
becomes so large that $SU(3)$ is no longer asymptotically free.  Hence models 
based on representations with $\ell >6$ require an intermediate breaking 
to $SU(n)$ with $n>3$, followed by a further breaking to $SU(3)$ accompanied 
by a reduction in the number of families.  
\bigskip
\centerline{\bf 5.~~The electroweak sector}

We turn next to an examination of how electroweak forces acting on the  
composite quarks and leptons arise in our model.  Far below the 
$SU(4)$ scale $\Lambda_H$, the gauge gluons that are present are 
the 9 horizontal gluons that mediate color $SU(3)$ and $U(1)$ forces and, 
we have argued, color singlet composites of the vertical gluons.  
We begin by showing that in the absence of a condensate breaking $Z_{12}$ 
to $Z_{6}$, the vertical composites  {\it cannot} account for the weak 
interactions.  To see this, consider a transition from a $d$ quark to 
a $u$ quark, which according to Eq.~(2) is, in rishon terms, a transition 
$\overline{T}\overline{V}\overline{V} \to VTT$, and by Eqs.~(3a), (3b), 
(4b) and (9) 
obeys the selection rules
$$\eqalign{
\Delta (n_T-n_V)=&\, 3\Delta (B-L) =  6 \Delta Q_{U(1)}=0\cr
\Delta (n_T+n_V)=&\,\Delta F =-\Delta n_{10}=6 ~~~.\cr
}\eqno(21a)$$
On the other hand, absorption of the color singlet composite of three 
vertical gluons 
in the $\overline{3}$ representation causes the transition $\overline{T}
VV \to \overline{V} TT$, and obeys the selection rules
$$\eqalign{
\Delta (n_T-n_V)=&\, 3 \Delta(B-L) = 6 \Delta Q_{U(1)} =6 \cr
\Delta (n_T+n_V)=&\,\Delta F =-\Delta n_{10}=0 ~~~.\cr
}\eqno(21b)$$
So evidently the selection rules for the weak interactions and for the 
composite of three vertical gluons are mismatched, and can be brought 
into correspondence only through the action of a Lorentz scalar 
(or pseudoscalar) color $SU(3)$ singlet electrically neutral 
condensate $\cal C$ obeying the selection rules, 
$$\eqalign{
\Delta (n_T-n_V)=&\, 3 \Delta(B-L) = 6 \Delta Q_{U(1)} =-6 \cr
\Delta (n_T+n_V)=&\,\Delta F =-\Delta n_{10}=6 ~~~,\cr
}\eqno(22a)$$
which, acting in conjunction with the absorption of a composite of 
three  vertical $\overline {3}$ representation gluons,  
converts the selection rules of Eq.~(21b) to those 
of Eq.~(21a).  Similarly, the absorption of a composite  of three vertical 
$3$ 
representation gluons requires the action of the conjugate condensate 
$\overline{\cal C}$   
obeying the corresponding selection rules, 
$$\eqalign{
\Delta (n_T-n_V)=&\, 3 \Delta(B-L) = 6 \Delta Q_{U(1)} =6 \cr
\Delta (n_T+n_V)=&\,\Delta F =-\Delta n_{10}=-6 ~~~,\cr
}\eqno(22b)$$
to produce the weak interaction transition $u \to d$.

Referring to the assignments of the $SU(3)$ components of the 
$\overline{10}_L$ state 
given in Eq.~(8), the following two possible condensates have the 
quantum numbers of Eq.~(22a), 
$$S_L^3T_L^3~,~~~V_L^6~~~,\eqno(23)$$
in both of which color and Lorentz indices are understood to be contracted to 
form,  
respectively, an $SU(3)$ singlet and a Lorentz scalar (or pseudoscalar).  
The following argument suggests that a condensate should indeed form 
in the $S_L^3T_L^3$ sector.  Consider first a pair $S_L T_L$, for which    
there is no $SU(3)$ binding force (the $S$ is a singlet) and for which 
the product of $U(1)$ charges in Eq.~(8), as well as of electric charges 
in Eq.~(9), is negative, so that the Abelian binding force is attractive 
on all length scales.  Since the interactions are attractive, these particles 
can form a loosely bound Lorentz scalar, color sextet, electrically 
neutral pair, and three such pairs can then bind by the $SU(3)$ color 
force to form a color singlet state.  Alternatively, we can  
argue that the three sextet $T_L$'s will 
bind to form a color singlet ``nucleus'' of $U(1)$ charge 1/2 and electric 
charge 1, that will then bind to three orbiting $S_L$'s each of $U(1)$ 
charge $-1/2$ and electric charge $-1/3$, to make an electrically neutral 
composite.  This argument also suggests that a $V_L^6$ condensate is 
unlikely, 
since the Abelian forces in this case are always repulsive or zero, and since 
$V_L^3$ can form a color singlet and the long range color force between 
two $V_L^3$ singlets vanishes.  Even in the absence 
of a $V_L^3$ condensate, the $S_L^3T_L^3$ condensate, which has the 
same quantum numbers, can cause effects such as neutrino-antineutrino 
mixing.  

To sum up the discussion thus far, we have argued that an explanation of the 
weak interactions requires, and the force and multiplet structure of our 
model makes 
it likely that there exists, a condensate of sixth degree in the fields 
of the 10 representation.  This condensate breaks the initial $Z_{12}$ 
discrete chiral symmetry to $Z_6$, and thus has exactly the structure needed 
in the analysis of Sec. 4 to convert the initial six massless families to   
three families.  In its role in allowing the exchange of vertical gluon 
composites to cause weak interaction transitions, the $S$ state functions  
as an analog of the ``doorway states'' [23] of nuclear physics.  As already 
discussed in the preceding section, we assume that the energy scale 
characterizing the $Z_6$  condensate is much smaller than the 
hypercolor scale at which preon binding occurs.  
Since the strong interactions play a role in the formation of the condensate, 
it is in fact reasonable to expect that the condensate 
scale, which helps set the electroweak scale in 
our model, should lie closer to the $SU(3)$ scale $\Lambda_{QCD}$ than to 
the hypercolor scale $\Lambda_H$.  A discussion of the relationship between 
$\Lambda_{QCD}$ and $\Lambda_H$ is given in Sec. 6.  

Let us now determine the algebraic structure of the resulting weak 
interactions, 
making throughout the assumption that the minimal algebra required by the 
quantum numbers is the one that occurs. 
Because the composites of three $\bar 3$ or three $3$ vertical gluons  
have electric charge $Q=\pm1$, 
their action on the fermionic composites takes place only within the 
doublet pairs $u,d$ or $e, \nu$  differing by one unit of electric charge. 
Hence the minimal algebraic structure containing their gauge charges is 
an $SU(2)$ of operators $W_{+}=W_1+iW_2, W_{-}=W_1-iW_2, W_3$, with $W_{+}$ 
the gauge 
charge associated with the composite formed from three $\overline{3}$ 
vertical gluons, with $W_{-}$ the gauge charge associated with 
the composite formed from three $3$ 
vertical gluons, and with 
$$\eqalign{
[W_{+},W_{-}]=&\,~2W_{3} \cr
[W_{3},W_{\pm}]=&\pm W_{\pm}  ~~~.\cr
}\eqno(24)$$
In Appendix A, we give an explicit calculation based on the $SU(4)$ gauge 
algebra, using the role of the $S$ as a doorway state, that shows that the 
algebra induced by the action of the vertical gluon composites, irrespective  
of the form of their $SU(3)$ wave function, has precisely the 
form of Eq.~(24).  Let us now take into account the effect on this algebra 
of the presence of a broken $Z_{12}$ symmetry.  If we imagine switching off 
the condensate, the charges associated with the vertical gluons still obey 
Eq.~(24), although they no longer cause weak interaction transitions.  
But now we must take into account the fact that vertical gluons 
can bind with preon pairs, and hence we can get bosonic composites with 
differing behaviors under the $Z_{12}$  chiral transformation of Eq.~(14).  
Because a real boson field $W_{1,2,3}$ must be self adjoint, it cannot 
transform under 
$Z_{12}$ with a complex phase; thus the only possibility allowed is 
that the vertical gluon composites  
pick up 6 preons in a charge zero, color singlet, Lorentz scalar state, 
producing a factor of $-1$ under the $Z_{12}$ transformation of Eq.~(14).
Hence before the condensate is turned on, we must have two distinct families 
$W_A^{e,o}$ of $SU(2)$ charges associated with three vertical gluon  
exchange, with the following behavior under the $Z_{12}$ transformations 
of Eq.~(14),
$$\eqalign{
W_A^e \to &\, W_A^e  \cr
W_A^o \to & \,W_A^o (-1)^k~~~, \cr
}\eqno(25)$$
which implies that both sets of charges are invariant under the 
$Z_6$ chiral subgroup that survives when the condensate is turned on.  
Identifying the charges $W^e$ with the $W$'s introduced above Eq.~(24), the 
minimal algebra with an $SU(2)$ structure 
in which Eqs.~(24) and (25) are both satisfied is evidently 
$$\eqalign{
[W_A^e,W_B^e]=&\,i\sum_C \epsilon_{ABC} W_C^e\cr 
[W_A^e,W_B^o]=&\,i\sum_C \epsilon_{ABC} W_C^o\cr 
[W_A^o,W_B^o]=&\,i\sum_C \epsilon_{ABC} W_C^e~~~,\cr 
}\eqno(26a)$$ 
which can be immediately diagonalized by forming the combinations 
$W_A^{e \pm o}$ to give 
$$\eqalign{
[W_A^{e+o},W_B^{e+o}]=&\,i\sum_C \epsilon_{ABC} W_C^{e+o}\cr 
[W_A^{e-o},W_B^{e-o}]=&\,i\sum_C \epsilon_{ABC} W_C^{e-o}~~~.\cr 
}\eqno(26b)$$
The algebra of Eq.~(26a) is isomorphic to the usual current algebra of 
vector and axial vector charges, and that of Eq.~(26b) to the diagonalization
of this algebra in terms of chiral charges.  Since the vector and axial 
vector charges give  the {\it only} isomorphic images of the algebras of 
Eq.~(26a,b) that can be formed from the composite fermions, the independent 
composite vector boson states $W_A^{e\pm o}$ {\it must} couple to the 
composite fermions with a chiral $SU(2)_L \times SU(2)_R$ gauge theory charge 
structure.  Finally, since we have seen in Sec.~2 that the $U(1)$ horizontal 
gluon couples to the composite fermions through the charge $B-L$, the 
low energy effective gauge theory of the composite fermions has the 
charge structure\footnote{*}{The charge $Q_{U(1)}$ can only commute with an  
$SU(2)$ obeying the selection rule $\Delta Q_{U(1)}=0$; hence Eq.~(21a) 
shows that $U(1)_{B-L}$ commutes with the physical $W$ states, which include 
the action of the condensate, whereas Eq.~(21b) shows that $U(1)_{B-L}$ does  
not commute with the composite of three vertical gluons uncoupled to the 
condensate.  We remark also that our argument for a chiral $SU(2)_{L} \times 
SU(2)_R$ structure based on a $Z_{12}$-induced doubling of the gauge algebra 
does not carry over to the vector-like $U(1)_{B-L}$ and color $SU(3)$
sectors, because the chiral currents which correspond to these are 
anomalous.}
$SU(2)_L \times SU(2)_R \times U(1)_{B-L}$.  

Note that this electroweak Lie group   
has rank 3, so that when the octet of horizontal gluons, which carry  
the color $SU(3)$ force, is included in the accounting, the low energy 
effective left-right symmetric theory has rank 5, that is, there is a   
set of 5 mutually commuting generators.  At first it may seem 
paradoxical that a fundamental $SU(4)$ theory of rank 3 could give rise 
to a low energy effective theory of rank 5.   Indeed, this would not be 
possible in a standard grand unification framework, where the low energy 
effective theory is always a subgroup of the full grand unification group, 
and so must have a rank not exceeding that of the unification group.  
However, in our model, only the $SU(3)_{\rm color} \times U(1)_{B-L}$ 
subgroup of the effective theory, which has rank 3, is constructed from  
generators that are a subgroup of the generator algebra of the unification 
group $SU(4)$, and this subgroup of the effective action obeys the usual   
rank restriction.   The remaining $SU(2)_L \times SU(2)_R$ group is 
constructed entirely from generators that are {\it composites} of the 
fundamental generators, and this permits 
the rank of the Lie algebra characterizing the low energy effective action   
to be enlarged.  
As shown in Appendix A, compositeness allows the  
the construction (when the chiral doubling is taken into account) of two  
additional generators that commute with each other, with the generator 
of $U(1)_{B-L}$, and with the color isospin and color hypercharge generators, 
leading in all to five mutually commuting generators and a rank 5 effective 
Lie algebra.  
                 
Let us now 
make the {\it assumption} that this charge structure is implemented by 
the dynamics as a local gauging; this is clearly a step that will have to 
be justified in future work.  We note, though, that recent results on  
supersymmetric models [3, 4] 
shows that there exist theories in which   
the low energy effective action contains composite local gauge degrees 
of freedom that differ qualitatively from those in the original Lagrangian.  
Generalizing from the example described in Sec. 5.4 of 
the lectures of Intriligator 
and Seiberg [4], 
one might infer that the rank of 
the dynamically generated gauge group should be smaller than the rank of the
fundamental gauge group, a restriction that is satisfied by our proposal   
since the rank of $SU(2)_L \times SU(2)_R$ is less than the rank of $SU(4)$.  
(Of course, from a limited class of examples, many generalizations are 
clearly possible.)  With the assumption of a local gauge dynamics, 
the low energy effective 
theory is an $SU(2)_L \times SU(2)_R \times U(1)_{B-L}$ gauge theory, 
which has a well-studied symmetry breaking pathway [8] leading to the 
$SU(2)_L \times U(1)_Y$ standard electroweak model.  A variant of the usual   
symmetry breaking discussion will be needed in our case, because as we have 
seen in Eq.~(22b) the condensate, which has nonzero $U(1)$ charge
$|Q_{U(1)}|=1$, already breaks the $U(1)$ gauge invariance.  

We emphasize that the argument for the development of an $SU(2)_L \times
SU(2)_R$ 
theory has not assumed parity violation, but only used the properties of 
the $Z_{12} \to Z_6$ symmetry breaking chain.  If the condensate responsible 
for this symmetry breaking is scalar, then the left and right gauge couplings 
$g_L$ and $g_R$ will be equal, and the theory is left-right symmetric.   
Parity violation then arises from the 
breaking of the left-right symmetric theory to the standard model.  An 
alternative possibility is that the condensate breaking the $Z_{12}$ symmetry 
is $P$, and $CP$, violating.  In this case the gauge couplings $g_L$ and 
$g_R$ will differ, and left-right symmetry is already violated at the level 
of  
the $SU(2)_L \times SU(2)_R$ theory.  This variant is also known [24] to have 
a symmetry breaking route to the standard model.  

In addition to the condensate ${\cal C}$ breaking $Z_{12}$ to $Z_6$, 
it is also possible 
that there is a second charge zero condensate of the form 
$S_LT_L\overline{V}_L^2$, with Lorentz and color indices contracted to 
form a scalar and singlet respectively, that breaks $Z_{12}$ to $Z_4$.  
The argument for three families in Sec.~4, and the analysis of electroweak 
structure 
in this section, can survive only if such an additional condensate is  
effectively a small perturbation on the condensate ${\cal C}$.  

We note in concluding this section that there are many 
questions connected with the 
symmetry breaking mechanism in our model that need further detailed study.  
Among them are: (i) We have argued that the two triplets, totaling 
6 vertical gluons, of the fundamental Lagrangian are replaced, in the 
effective Lagrangian acting on fermionic composites, by two triplets of 
$SU(2)$ gauge bosons, again totaling 6 gauge fields.  Is this correspondence 
of numbers a coincidence, or does it have deeper significance?  
To answer this question, one 
will need a generalization of the standard analysis [25] of effective 
Lagrangians 
to the case in which the transformation between phenomenological field 
variables $\phi$ and fundamental field variables $\chi$ has, instead of 
the regular form $\phi=\chi F[\chi]~,~~F[0]=1$, the singular form 
$\phi=\chi^3 F[\chi]~,~~F[0]=1$.    (ii)  We have only employed symmetry 
breaking mechanisms that break {\it discrete} global invariances, never 
continuous global invariances.  Are there corresponding Goldstone bosons?  
Since the condensate breaks the local  $SU(4)$ gauge invariance, and since 
the breaking of local gauge invariances is possible only after gauge 
fixing [26], it would appear that any associated Goldstone bosons should be 
gauge variant, and therefore may not be physical.  
Even if physical, by the analysis of [25] they will consist of a color 
triplet and antitriplet of scalars (analogous to the vertical gluons) 
corresponding  
to the coset manifold $SU(4)/SU(3)$, and by our binding postulate should 
bind to form color singlet composites.  Can these composites, in analogy 
with conventional technicolor scenarios [8, 27], 
play a role in the electroweak 
Higgs sector?  These are  subtle issues that need further study.  
(iii)  As we have already noted, the condensate has 
integer $U(1)$ charge and therefore partially breaks the original $U(1)$ 
gauge 
invariance, so that $Q_{U(1)}$ is only conserved modulo an integer and 
$B-L$ is only conserved modulo 2.  At the same time, the condensate is 
electrically neutral and so does not break conservation of the electric 
charge $Q$.  Thus our model is consistent with the emergence, after all 
symmetry breakings, of an unbroken electromagnetic $U(1)$ gauge field, but 
the detailed route by which this happens has to be established.  
 
\bigskip
\centerline{\bf 6.~~The renormalization group and the gauge hierarchy}

According                                                                      
                                                                               
                                  to the picture developed in the preceding 
sections, the gauge 
forces associated with $SU(4)$ are responsible both for preonic binding at a 
high scale $\Lambda_H$, and for the electroweak forces and the strong QCD 
force at much lower scales $\Lambda_{EW}$ and $\Lambda_{QCD}$.  What 
can be said about the relation between these scales?  An analysis of the 
relation between $\Lambda_{EW}$ and $\Lambda_{QCD}$, and of the value of  
the Weinberg angle, will require a detailed understanding of the dynamics 
of the condensate and its effect on the $U(1)$ couplings, and will not be 
attempted here.  However, since the condensate is a color singlet it has no 
effect on the evolution of the $SU(3)$ coupling, allowing us to consider 
the relation between $\Lambda_H$ and $\Lambda_{QCD}$ as a simpler problem 
isolated from electroweak complications.  

There are then three regimes to consider, as shown in Fig. 1, 
consisting of two branches of the   
running coupling, separated by a nonperturbative 
strong coupling regime in which a 
running coupling is insufficient to describe the dynamics.\footnote{*}  
{There are actually four regimes to be considered, since the running of the 
$SU(3)$ coupling speeds up below the scale where the three heavy partners  
of the quarks in the three families are frozen out;  we shall assume that 
this scale is much closer to $\Lambda_{QCD}$ than to $\Lambda_H$, and 
ignore this complication in the following discussion.}  
For energies well above the 
scale $\Lambda_H$, branch 1 of the 
running coupling, describing the interaction  
of the preons, is given by the $SU(4)$ formula 
$$g^2_{SU(4)}(\mu^2)={1 \over b_{SU(4)} \log(\mu^2/\Lambda_H^2) }~~~,
\eqno(27a)$$
while for energies well below $\Lambda_H$ and well above $\Lambda_{QCD}$, 
branch 2 of the running coupling, describing the residual color 
interactions of composite quarks, is given by the $SU(3)$ formula 
$$g^2_{SU(3)}(\mu^2)={1 \over b_{SU(3)} \log(\mu^2/\Lambda_{QCD}^2) }~~~.
\eqno(27b)$$
For energies in the vicinity of $\Lambda_H$, neither of these formulas 
applies, since the $SU(4)$ preonic coupling is strong, leading to the 
binding of both preons and vertical gluons on a 
length scale $\Lambda_H^{-1}$.
Tracing the dynamics across this strong coupling nonperturbative 
region separating the 
two branches is a complicated 
but in principle calculable problem, the result of which 
will be a ``connection 
formula'' fixing the relation between the scales $\Lambda_H$ and 
$\Lambda_{QCD}$ characterizing the two branches of the running coupling.  
The result of this calculation, we shall now show, can 
always be expressed as the value of the coupling $g_*^2$ at the point 
$\mu^2$ where Eq.~(27a) 
and the high energy extrapolation of Eq.~(27b) intersect, i.e., where
$$g_*^2\equiv  g^2_{SU(4)}(\mu^2)=g^2_{SU(3)}(\mu^2)~~~.\eqno(28)$$
Such an intersection always exists provided that $b_{SU(4)}$ is larger 
than $b_{SU(3)}$, in other words,  
provided the $SU(4)$ coupling is running faster 
than the $SU(3)$ coupling, since then the ratio 
$${g^2_{SU(3)}(\mu^2) \over g^2_{SU(4)}(\mu^2)} \eqno(29a)$$ 
is smaller than unity for $\mu^2$ just above $\Lambda_H$  but 
approaches the value 
$${b_{SU(4)} \over b_{SU(3)} }~~~, \eqno(29b)$$
which is larger than unity, as $\mu^2 \to \infty$.  In terms of the 
coupling $g_*^2$ at the intersection, 
the relation between the scales $\Lambda_H$ and 
$\Lambda_{QCD}$ takes the form 
$${\Lambda_H^2 \over \Lambda_{QCD}^2 } = 
\exp[(b_{SU(3)}^{-1}-b_{SU(4)}^{-1})/g_*^2]~~~.\eqno(30)$$

The values of the renormalization group beta function coefficients 
$b_{SU(3),SU(4)}$ can be computed from the formula [9] 
$$b=-{\mu \over g^3}{dg \over d \mu} =
{1 \over 32 \pi^2}\left({11 \over 3} \ell({\rm vector})-{4\over 3}
\ell({\rm Dirac~fermion}) -{1 \over 6} \ell({\rm real~spinless})\right) 
~~~,\eqno(31)$$
where the $\ell$ values are the sums of the Slansky indices for the 
representations 
in which the particles lie.\footnote{*}  
{In Eq.~(31) the familiar constant $16 \pi^2$ has been 
replaced by  $32 \pi^2$, to compensate for the fact that Slansky's index 
$\ell$ is normalized to be always an integer, so that in Slansky's tables 
$\ell({\rm fundamental})=1$ and $\ell({\rm adjoint})=2N$ for the group 
$SU(N)$.  Equation~(31) agrees with the standard result $b=(16 \pi^2)^{-1}
(11-2N_f/3)$ for $N_f$ flavor QCD.}
To make a preliminary estimate, we defer the issue of dealing with possible  
fundamental or composite scalars in our model, and thus simply ignore 
a possible spin zero contribution 
to Eq.~(31).  Then in the formula for $b_{SU(4)}$ we have  
$\ell({\rm vector})=\ell(15)=8$ and $\ell({\rm Dirac~fermion})=\ell(10)=6$, 
while in the formula for $b_{SU(3)}$ we have $\ell({\rm vector})=\ell(8)=6$ 
and $\ell({\rm Dirac~fermion})=12\ell(3)=12$, where we have taken 
into account the 
fact that the preons bind to give 6 families, each containing 2 quark 
triplets.  Substituting these values into Eq.~(31) and then evaluating 
Eq.~(30), we find 
$${\Lambda_H^2 \over \Lambda_{QCD}^2 } = 
\exp[23 \pi^2/(6 g_*^2)]=\exp(37.8/g_*^2)~~~.\eqno(32)$$
The relatively large exponent in Eq.~(32) is a result of the small ratio  
$b_{SU(3)}/b_{SU(4)}=9/32$, expressing the fact that the $SU(3)$ 
coupling runs much more slowly than the $SU(4)$ coupling as a result of the 
large number of composites created by the binding of the preons,  and so 
in this respect the behavior of our model is reminiscent of the behavior 
of ``walking technicolor'' [27] theories.  Equation (32) can give a large 
hierarchy ratio for values of $g_*$ not much smaller than unity; for example,  

for $g_*=0.7$ one finds $\Lambda_H \sim 10^{16} {\rm GeV}$, and hence 
a large hierarchy ratio is natural in our model.  

For our model to be compatible with experimental limits on proton decay, 
it is crucial that the hierarchy ratio be very large, and 
that the vertical gluons, as well as the preons, be confined 
within a radius $\Lambda_H^{-1}$.  Exchange of a single unbound 
vertical gluon 
can lead to the reaction $u+u \to \bar d + e^{+}$, and hence to proton decay.  

With confined vertical gluons and preons, the reaction amplitude 
is proportional to the cross sectional area of a $u$ quark,  
and hence the rate varies as $\Lambda_H^{-4}$.  This dependence 
of the proton decay rate on the high mass scale 
is compatible [28] with experimental bounds, for values 
of $\Lambda_H$ well below the Planck mass.  As long as $\Lambda_H$ is big 
enough for proton decay to be below current experimental limits, other 
effective four fermion (i.e., dimension 6) 
interactions resulting from the composite structure 
of quarks and leptons will automatically be much smaller than the current 
limits on such effective action terms.  
\vfill
\eject
\centerline{\bf 7.~~Discussion:  possible supergravity embeddings}
\centerline{\bf and phenomenological implications}
We turn finally to the issue of the embedding of our model in a larger 
structure, including gravitation, and to a brief discussion 
of phenomenological implications.  
We first remark that if our model is 
augmented by an $SU(4)$ 6 representation of Majorana fermions, the basic 
mechanisms discussed above can still work.  
The quadrality of the 6 is also 2, 
and so the $SU(4)$ forces remain frustrated in the fermionic sector.  If 
the 6 representation develops its own $Z_2$ condensate at a high mass scale, 
it acts as an inert spectator in the $Z_{12}$ chiral symmetry breaking chain 
for composites formed within the  10 representation, 
although it could make a contribution to the masses of these composites in 
the 
final stages of chiral symmetry breaking.  

Let us now turn to consider gravity, or more specifically, supergravity.  
The existence of extended supergravity multiplets at one time briefly 
raised the hope that all matter fields, and gravitation, could be unified 
within one extended graviton multiplet.  This hope could not be realized, 
however, in the framework of grand unification models or direct product 
hypercolor composite models, because the symmetry groups involved are too 
large to fit within the largest $(N=8)$ extended graviton multiplet [10].  
An 
attractive feature of the model presented here is that this objection 
no longer applies: as already noted above,       
the $SU(4)$ symmetry group employed, and the required  
15 spin one gauge gluons and 10 Dirac fermions  (equivalent to 20 Majorana 
or Weyl fermions) correspond to representations appearing in the   
$N=6$ supergraviton multiplet [10, 12] after  
reflection to insure 
$CPT$ invariance. (The $CPT$ invariant structure has an additional 6 
Majorana fermions, that we noted are acceptable additions to the model.)  
Hence if the model given here 
proves viable, it becomes an important question to construct a  
supergravity (and perhaps also a conformal supergravity [29]) field theory 
Lagrangian 
with the requisite gauge gluon and fermion representation structures.  
Such a theory 
would provide the possibility of a truly elegant unification of all of the 
forces.  We note, finally, that any extended supergravity embedding of our 
model will contain fundamental scalars, with a representation structure 
that is dependent on the structural details of the extension.  It is for 
this reason that we have not attempted at this stage an exploration of the 
scalar structure of our model; without solving the problem of classifying 
the possible supergravity embeddings, there appear to be many possibilities 
for both fundamental and composite scalars consistent with the vector 
and spinor dynamics that we have outlined.   

The principal phenomenological implication of our model is that for each 
of the three standard model 
families of quarks and leptons, there is a corresponding 
heavy family with the same quantum numbers ({\it not} a mirror family),  
some members of which could have 
masses as low as those characterizing the third standard model family.  
We must first ask whether the presence of such extra families is  
compatible with experiment?  
Under the simplifying assumption that the effect of these heavy fermions  
on electroweak radiative corrections can be parameterized solely through 
the $S$ parameter [30], the existence of 3 heavy families (6 heavy flavors)
leads\footnote{*}{I am indebted to R. N. Mohapatra for this observation.}  
to a contribution to $S$ of ${2 \over \pi}$, which is only marginally 
compatible with the LEP data as reported in [30].  However, two caveats 
are relevant here.  The first is that in our model 
the heavy states can mix with their standard model counterparts, 
and consequently [30] the implications of these states 
for electroweak physics cannot be fully parameterized 
within the standard $S,T,U$ phenomenological framework.  
The second is that scalar particles can lead to 
negative contributions 
to $S$ (see, e.g., [30, 31]), and a supergravity embedding of our model will 
contain many scalars.  Thus, electoweak radiative corrections do not at this   
 
point make a decisive statement about the viability of our prediction  
of heavy family partners.  
We remark that in an $N=6$ supersymmetry embedding of our model, 
these heavy partners could be the first signal for supersymmetry, contrasting  

with the prediction of an ``s'' partner for each  
standard model particle by $N=1$ supersymmetry extensions of the standard 
model.

Recently, the H1 and ZEUS collaborations at HERA have reported [32] 
an excess of positron-jet events at large $Q^2$, and so it is natural to 
ask if these events might signal the discovery of a heavy family partner 
$E^+$ of the positron $e^+$.  As formulated up to this point, our model 
cannot account for the HERA events; the reason is that if color $SU(3)$ 
remains an exact 
symmetry, and if color neutralization is assumed to be instantaneous, the 
$E^+$ would be produced as a color singlet, and its dominant decay mode 
would be the electromagnetic decay $E^+ \rightarrow e^++\gamma$, which does 
not correspond to the $e^+$ plus jet signature reported by the HERA groups.  
However, the production and decay modes of the $E^+$ in our model depend 
strongly on the details of the surviving unbroken symmetries, and of 
color neutralization. 
                      
Suppose, as suggested by Slansky, Goldman, and Shaw [33], 
that color $SU(3)$ were weakly broken to ``glow'' 
$SO(3)$, while maintaining triality conservation modulo 3, 
with color gluons not in the 
$SO(3)$ subgroup acquiring very small masses, and with the threshold for 
excitation of free color dropping from infinity to a finite value 
of order 200-300 GeV. [It makes sense to talk  
about a free color threshold in this context because under $SU(3) \supset 
SO(3)$, all states of $SU(3)$ decompose into diality zero integer $L$ 
representations of SO(3); half-integer, nonzero diality representations of 
the covering group $SU(2)$, which are confined, are never encountered.]  
The possibility of the breaking of color to glow can naturally be 
incorporated into our model, for two reasons.  First, the state 
classification of Secs. 3 and 4 uses only the conservation of triality 
modulo 3, but not the details of the particular $SU(3)$ representations 
corresponding to each triality sector, and so a triality preserving breaking 
of $SU(3)$ to $SO(3)$ leaves this classification, and in particular the 
distinction between leptons and quarks, intact.  Second, to achieve a 
triality preserving breaking of $SU(3)$ to $SO(3)$, the smallest 
non-singlet $SU(3)$ representation with nonzero vacuum expectation must [33] 
be the 27, since this is the smallest triality zero representation of $SU(3)$ 
that contains an $SO(3)$ singlet.  But since the postulated condensate 
$S_L^3T_L^3$ of our model has an $SU(3)$ tensor product structure 
corresponding to the symmetric part of 
$6 \times 6 \times 6$, which contains the 27, it is consistent with the 
framework developed above to additionally postulate that the condensate 
has a small component in the 27 representation of color $SU(3)$ as well as 
a dominant color singlet component.  The effect of a breaking of color 
$SU(3)$ to glow $SO(3)$, with a free color excitation threshold of order 
200-300 GeV, would be to leave the standard model leptons as color singlets, 
but to permit their heavy counterparts to carry admixtures of color 
non-singlet states.  Similarly, the heavy counterparts of the quarks would 
carry admixtures of triality $\pm1$ states other than the states 
$3,\overline 3$, and electroweak anomaly cancellation would then 
give a sum rule relating the quark color state 
admixtures to those in the lepton states.  
In this scenario, the $E^+$ could carry a color octet component.  It  
would be produced by positron gluon collision, and its dominant 
decay mode would be single color gluon emission $E^+ \rightarrow e^+ 
+g$, which would appear as a positron jet final state, corresponding to 
a ``leptogluon'' interpretation of the signature observed at HERA.  As 
discussed in a recent phenomenological analysis of Akama, Katasuura, and 
Terazawa [34], such an ``excited positron'' interpretation is consistent with 
the HERA data and with limits from other accelerator experiments.  

An alternative scenario, which does not require color $SU(3)$ to be broken, 
is simply to assume that color neutralization does not fully take place in 
the hard processes involved in $E^+$ production and its subsequent decay.  
Recall that in our model, the $e^+$ and $E^+$ are both $TTT$ three preon 
bound states, with an $SU(3)$ wave function (before color neutralization) 
corresponding to the mixed symmetry part of $6_L \times 6_L \times 6_R$,  
which has the Clebsch series 
$8 + 10 + \overline 10 +...$ and contains no color singlet.  Our postulate 
of Sec. 3 is that color neutralization occurs by picking up color gluons 
from the vacuum until the $SU(3)$ state with lowest Casimir is attained, 
so that only the $SU(3)$ triality plays a role in enumerating the possible 
states.  However, in very hard processes, characterized by momentum transfers 
much larger than the QCD scale, it is possible that this color neutralization 
could be incomplete, and that the $E^+$ would then behave as a state with 
the color wave function suggested by the bare preon Clebsch series.  Again, 
as in the color breaking scenario, this would permit the production of 
the $E^+$ by positron gluon collision, and its subsequent rapid decay into 
a positron and a gluon jet.  

In summary, we suggest that the production and decay of the excess HERA 
events, interpreted as leptogluons, 
could be accounted for in our model when augmented by either the 
assumption that the $Z_6$ condensate that breaks $SU(4)$ to color $SU(3)$ 
contains a small component that further breaks color $SU(3)$ to glow $SO(3)$, 
or by the assumption that color symmetry remains exact but that color 
neutralization is incomplete in hard processes.  On the other hand, a 
leptoquark interpretation of the HERA events is not apparent in our model; 
composite vector leptoquarks would be expected to have masses 
near $\Lambda_H$, since there is no chiral or gauge 
symmetry argument for them to have small masses.  Assessing the possibility 
of scalar leptoquarks will require further study of the related problems 
of supergravity embeddings and the scalar sector of our model.

\bigskip
\centerline{\bf Acknowledgments}
I wish to acknowledge fruitful interactions with T. Applequist, W. Bardeen, 
P. Binetruy, S. Coleman, A. Faraggi, E. D'Hoker, K. Dienes, E. Eichten, 
D. Fairlie, D. Freedman, H. Georgi, L. Horwitz, 
K. Intriligator, S. Leibler, A. Millard, J. March-Russell, 
R. Mohapatra, A. Nelson, J. Nuyts, D. Politzer, 
R. Shrock, N. Seiberg, R. Slansky, E. Simmons, M. Strassler, F. Wilczek, 
E. Weinberg, E. Witten, and B. Zumino.  This work was begun during a visit to 
the Aspen 
Center for Physics, and was supported in part by the Department of Energy 
under
Grant \#DE--FG02--90ER40542.  
\medskip
{\it Added note.} After this manuscript was initially 
posted on the theory Bulletin 
Board, I received 
an interesting email from D. Fairlie and J. Nuyts who pointed out that 
the model given here, in which the fermions are in the 10 of $SU(4)$, 
fits into the general group theoretic framework of Fairlie, Nuyts and 
Taormina 
[35].   This paper (see its Appendix A) 
showed that a large class of preonic models constructed from 
fundamental preons  must have exotic charge 1/6 fermion 
states, except in a rishon type model in which the rishons are composed 
of symmetrized pairs of preons (as in the 10 of $SU(4)$ used here, 
which is the symmetrical tensor product $4 \times 4$).

\vfill
\eject
\centerline{\bf Appendix A. 
~~The three vertical gluon composite gauge algebra}

We compute here the gauge algebra corresponding to the action of a three 
vertical gluon composite on a preon bound state.  Since the gluons couple 
to the preons through the representation matrices for the 10 representation 
of $SU(4)$, we must first construct these matrices.  We do this by 
starting with the representation matrices $\Lambda_A$ for the fundamental 
4 representation (an explicit representation for them will be given shortly), 
and using the fact that since $4\times 4 = 6 +10$, the 10 representation 
is the symmetric part of the tensor product of two 4's.  Let $(ab)$ denote 
an index pair with $a,b=1,...,4$, with the parentheses implying 
symmetrization, so that there are only 10 distinct values of $(ab)$ when 
$(ab)$ and $(ba)$ are treated as equivalent.   Then we can use $(ab)$ 
as a label for the 16 rows and 16 columns 
of the representation matrices $M^{(ab)}_{A (\bar a \bar b)}$ for the 10 
representation.  A simple computation using the group transformation 
law for a tensor product then gives
$$M^{(ab)}_{A (\bar a \bar b)}={1 \over 2}
\left(\Lambda^a_{A \bar a} \delta^b_{\bar b}
     +\Lambda^b_{A \bar b} \delta^a_{\bar a}
     +\Lambda^b_{A \bar a} \delta^a_{\bar b} 
     +\Lambda^a_{A \bar b} \delta^b_{\bar a}  \right)  ~~~,\eqno(A1)$$
with $\delta^a_b=\delta_{ab}$ the Kronecker delta.

As discussed in the text, the 10 representation of $SU(4)$ 
contains a 1, a 3, and a 6 of $SU(3)$; let us order these so that the 
label $(44)$ denotes the 1, the label $(4a),~a=1,2,3$ denotes the 3, and 
the label $(ab),~a,b=1,2,3$ denotes the 6.  We have seen that the 
weak interactions require the intervention of a condensate $S_L^3T_L^3$, 
where $\overline{S},\overline{V},\overline{T}$ denote respectively the 
$SU(3)$ states $1,3,6$.  Thus (ignoring the possibility of 
neutrino-antineutrino mixing) the only vertical gluon transitions between 
the preons relevant for the weak interactions are those between the 
$\overline{V}$ and the $\overline{S}$, described by the submatrix of  
Eq.~(A1) with one index pair equal to $(44)$ and one index pair equal to 
$(4a),~a=1,2,3$, and with the index $A$ corresponding to the triplet or   
antitriplet of vertical gluons.  

At this point in the calculation it is convenient to introduce a specific 
representation for the $SU(4)$ fundamental representation matrices 
$\Lambda_A$.  Let $\lambda_A, A=1,...,8$ be the standard Gell-Mann matrices 
for $SU(3)$; then we take the first 8 $SU(4)$ matrices to be 
$$\Lambda_A={\rm diag}(\lambda_A,0)~~~,\eqno(A2)$$
where we use the notation ${\rm diag}(\alpha, \beta)$ to indicate a $4 \times 
4$ 
block diagonal matrix with a $3 \times 3$ diagonal block $\alpha$ 
and a $1 \times 1$ diagonal block $\beta$.  The remaining 7 $SU(4)$ matrices 
consist of the $U(1)$ generator 
$$\Lambda_{15}=6^{-1/2} {\rm diag}(1,-3) ~~~,\eqno(A3)$$
and the six generators $\Lambda_{9,...,14}$ for the vertical gluons.  
It is convenient to 
write the latter in the form of an $SU(3)$ triplet and an $SU(3)$ antitriplet  

of non-self-adjoint raising and 
lowering operators, $\tau_{\pm \, k},~k=1,2,3$, defined by  
$$\eqalign{
(\tau_{+ \, k})^a_{\bar a}=&\,\delta^a_k \delta^4_{\bar a} \cr
(\tau_{- \, k})^a_{\bar a}=&\,\delta^a_4 \delta^k_{\bar a} ~~~,\cr
}\eqno(A4)$$
which obey the algebra
$$\eqalign{
\tau_{+\,k}\tau_{+\,\ell}=&\,\tau_{-\,k}\tau_{-\,\ell}=0  \cr
\tau_{+\,k}\tau_{-\,\ell}=&\,{\rm diag}(D_{k \ell},0) \cr
\tau_{-\,k}\tau_{+\,\ell}=&\,{\rm diag}(0,\delta_{k\ell}) ~~~,\cr
}\eqno(A5a)$$
where $D_{k\ell}$ is the $3 \times 3$ matrix with matrix elements 
$$(D_{k\ell})^a_{\bar a}= \delta^a_k \delta_{\bar a \ell}~~~,\eqno(A5b)$$
together with
$$\tau_{+ \, k} {\rm diag}(0,1)=\tau_{+ \, k}~,~~~
{\rm diag}(0,1) \tau_{- \, k}=\tau_{- \, k}~~~.\eqno(A5c)$$
If we now write the 10 representation matrices of Eq.~(A1) for the 6 
vertical gluons in corresponding non-self-adjoint raising and lowering 
operator form, we get 
$$M^{(ab)}_{\pm\,k (\bar a \bar b)}={1 \over 2}
\left(\tau^a_{\pm\,k \,\bar a} \delta^b_{\bar b}
     +\tau^b_{\pm\,k \,\bar b} \delta^a_{\bar a}
     +\tau^b_{\pm\,k \,\bar a} \delta^a_{\bar b} 
     +\tau^a_{\pm\,k \,\bar b} \delta^b_{\bar a}  \right)  ~~~.\eqno(A6)$$
>From this, we find for the submatrix acting on the $SU(3)$ 1 and 3 
states $\overline{S}$ and $\overline{V}$,   
$$\eqalign{
M^{(4a)}_{\pm\,k (4b)}=&\,M^{(44)}_{\pm\,k (44)}=0 \cr
M^{(44)}_{+\,k (4a)}=&\, M^{(4a)}_{-\,k (44)}=0  \cr
M^{(4a)}_{+\,k (44)}=&\,(\tau_{+\,k})^a_4  \cr
M^{(44)}_{-\,k (4a)}=&\,(\tau_{-\,k})^4_a   ~~~,\cr
}\eqno(A7)$$
with $a,b=1,2,3$ only.  
Thus this submatrix  has the same structure as the corresponding 4 
representation matrix $\tau_{\pm\,k}$ 
acting on a 4 of  $SU(4)$  constructed from the 
$\overline{S}$ and $\overline{V}$ states, and 
therefore to study the algebra of the three 
vertical gluon composites as it acts in the 
$\overline{S}\overline{V}$ subspace, it suffices 
to study this algebra using the 4 representation matrices of Eq.~(A4).

This calculation is relatively straightforward.  The charge operators 
$U_{\pm}$, which describe the action on 4 representation  preon triples 
with charge matrices $\tau^{(1)},~ \tau^{(2)},~\tau^{(3)}$, 
of the triality 0 composites 
formed from the triplet and antitriplet of vertical gluons, are
$$U_{\pm}=\sum_{kmn} C_{kmn} \tau^{(1)}_{\pm\,k}\tau^{(2)}_{\pm\,m}
\tau^{(3)}_{\pm\,n}~~~,\eqno(A8)$$
with $C_{kmn}$ a tensor determined by the internal structure of the 
three gluon composite.  From the algebraic properties of $\tau$ matrices 
in Eq.~(A5a-c), we see that 
$$\eqalign{
U_{+}^2=&\,U_{-}^2=0 \cr
U_{-}U_{+}=&\,KP\cr
U_{+}P=&\,U_{+} \cr
PU_{-}=&\,U_{-}~~~,\cr
}\eqno(A9a)$$
with $K$ the constant
$$K=\sum_{kmn} C_{kmn}^2~~~,\eqno(A9b)$$ 
and with $P$ the projector
$$P= {\rm diag}(0,1)^{(1)} {\rm diag}(0,1)^{(2)}{\rm diag}(0,1)^{(3)} 
~~~.\eqno(A9c)$$
So defining charge operators $W_{\pm}$ and $W_3$ by 
$$W_{\pm}=K^{-1/2} U_{\pm}~,~~W_3={1 \over 2K}[U_{+},U_{-}]~~~,\eqno(A10a)$$
we see that 
$$\eqalign{
[W_{+},W_{-}]=&2W_3 \cr
[W_3,W_{+}]=&\,{1 \over 2 K^{3/2} }(U_{+}U_{-}U_{+}-U_{-}U_{+}U_{+}
-U_{+}U_{+}U_{-}+U_{+}U_{-}U_{+})\cr
=&\,{1 \over 2 K^{3/2}} 2 U_{+}U_{-}U_{+}=K^{-1/2} U_{+}=W_{+}\cr
[W_3,W_{-}]=&\,{1 \over 2 K^{3/2} }(U_{+}U_{-}U_{-}-U_{-}U_{+}U_{-}
-U_{-}U_{+}U_{-}+U_{-}U_{-}U_{+})\cr
=&\,{-1 \over 2 K^{3/2}} 2 U_{-}U_{+}U_{-}=-K^{-1/2} U_{-}=-W_{-}~~~.\cr
}\eqno(A10b)$$
Thus the $W$ charges obey the $SU(2)$ algebra of Eq.~(24) of the text, 
irrespective of the detailed structure of the internal wave function 
$C_{kmn}$.

In the special case in which the tensor $C_{kmn}$ is the $SU(3)$ structure 
constant $f_{kmn}$, the composites $U_{\pm}$ are color singlets and 
commute with the whole horizontal $SU(3)$ Lie algebra, without requiring 
the addition of horizontal gluons to achieve color neutralization.  
Therefore, the commutator $[U_+,U_-]$ commutes with the $SU(3)$ Lie 
algebra; since this commutator carries $U(1)$ charge zero it also commutes 
with the $U(1)$ generator, and so a rank 4 set of generators is given  
by the $U(1)$ 
generator, the third component of color isospin and the color hypercharge, 
and the commutator $[U_+,U_-]$.  This is an explicit example showing that 
when composite structures in the group generators are allowed, one can 
get effective Lie algebras with {\it higher} rank than that of the 
fundamental Lie algebra from which they are formed.  [Since the left-right 
symmetric theory $SU(3) \times SU(2)_L \times SU(2)_R \times U(1)_{B-L}$ 
is actually of rank 5, we need one more mutually commuting generator, and 
we argue in Eqs.~(24-26) of the text that this comes from the breaking 
of the discrete $Z_{12}$ chiral symmetry to $Z_6$.]
\bigskip
\centerline{\bf Appendix B. ~~Mixed preon-antipreon fermionic states}

As noted at the end of Sec.~3 of the text, in addition to the three 
preon fermionic composites of Eq.~(2), our model also permits the 
fractionally charged, triality zero, mixed preon-antipreon fermionic states 
$$\eqalign{
\ell_U =&TT\overline{V}  \cr
\overline{\ell}_D =&T\overline{V}\overline{V}~~~, \cr
}\eqno(B1)$$
with respective charges $Q=2/3$ and $Q=1/3$, 
together with their corresponding antiparticles.  [In the Harari-Shupe 
scheme [6], the absence of these states is enforced through an ad hoc ``no 
mixing''rule; in the Harari-Seiberg model [5], this rule is implemented 
through 
the hypercolor triality assignments of the rishons, which prevents 
the states of Eq.~(B1) from being hypercolor singlets.]  For these states,   
we readily find that the analogs of the counting relations of Eq.~(3b) are 
$$\eqalign{
{1 \over 3}(n_T-n_V)=&n_{\ell_U}-n_{\ell_D} \cr
{1 \over 3}(n_T+n_V)=&{1 \over 3}(n_{\ell_U}+n_{\ell_D})~~~. \cr
}\eqno(B2)$$
A transition $\ell_D \to \ell_U$ evidently obeys the selection rules 
$$\eqalign{
\Delta(n_T-n_V)=&6 \cr
\Delta(n_T+n_V)=&0 ~~~,\cr
}\eqno(B3)$$
which agree with those of Eq.~(21b) for a transition induced by a three 
vertical gluon composite {\it without} the action of the condensate $\cal C$.  

Also, we see that for the states of Eq.~(B1), the charge $n_T-n_V$, 
which acted (with the condensate) as the electroweak $U(1)$  for the 
composites of Eqs.~(2) and (3b), now acts as the third component of  
an electroweak $SU(2)$.  Conversely, for the states of Eq.~(B1), 
the charge $(n_T+n_V)/3$, which acted as the third component of the 
electroweak 
$SU(2)$ for the composites of Eqs.~(2) and (3b), now acts as an 
electroweak $U(1)$.  Thus the electroweak interactions for the composites  
of Eq.~(B1) have a structure incompatible with those for the standard 
model composites of Eq.~(2).  This point is further underscored if we 
add the helicity indices $L,R$ to the electroweak groups and look at the 
corresponding anomaly structure.  The ordinary family particles do 
not produce an anomaly in the $SU(2)$ current with chiral charge $n_T + n_V$, 
but do produce an anomaly in the $U(1)$ current with chiral charge $n_T-n_V$.  

Conversely, the states of Eq.~(B1) produce 
no anomaly in the $SU(2)$ current with chiral charge $n_T-n_V$, but 
do produce an anomaly in the $U(1)$ current with chiral charge $(n_T+n_V)/3$.

These facts mean that the low energy effective action of our model 
cannot simultaneously contain both the standard model particles of Eq.~(2), 
and the fractionally charged particles of Eq.~(B1), along with their 
electroweak gaugings.  We interpret this to mean that our model can 
exist in two phases.  In one phase, in which there is no $S_L^3T_L^3$ 
condensate, the low energy spectrum consists of the fractionally 
charged particles of Eq.~(B1) and their strong and electroweak gauge bosons.   

In the second phase, which we are assuming to be the physically realized  
one, there is an $S_L^3T_L^3$ condensate and the low energy spectrum consists 
of the standard model fermions of Eq.~(2) and their strong and electroweak  
gauge bosons.  In the assumed physical phase, there is a mechanism for 
giving the fractionally charged states of Eq.~(B1) masses at the high 
scale $\Lambda_H$, since self energy graphs in which these states emit 
a three vertical gluon composite with chiral charge $n_T-n_V$ (without 
intervention of the condensate) will 
receive a divergent contribution from back to back insertions 
on the composite gluon propagator  of anomalous 
triangle graphs containing the standard model fermions.  Thus, according 
to the interpretation suggested here, the discussion of Secs.~4 and 5 of the 
text extends to show that the ``no mixing'' rule also emerges from the  
chiral symmetry structure of our model.  
\bigskip
\centerline{\bf Appendix C.~~Does the Coleman-Mandula Theorem}   
\centerline{\bf Apply to Discrete Chiral Symmetries?}
In the chiral symmetry analysis of Sec.~4, we made the technical assumption 
that we only had to consider mass terms {\it within} groups of states that 
are allowed to couple through the kinetic energy, and hence are classified 
as members of the same family.  To see why this assumption is needed,  
consider the inter-family off-diagonal mass term 
$$\Psi_{MR}^{\dagger}\gamma_0 \Psi_{NL}~+~{\rm adjoint}~~~,\eqno(C1a)$$
which is invariant under the $Z_{12}$ chiral transformation of 
Eqs.~(14a, b) when 
$$(2M+1+2N+1)k/12=(M+N+1)k/6~~~\eqno(C1b)$$ 
is an integer for all $k$, a condition that is satisfied for the 
combinations $(M,N)=$(0,5), (1,4), (2,3).  Similarly, a discrete $Z_6$  
chiral invariance allows off-diagonal mass couplings between monomial pairs 
$(M,N)=(0,2),(1,1)$.  In either of these cases, diagonalizing the mass term 
leads to mass eigenstates that are not discrete chiral symmetry eigenstates,   

because the kinetic energy term for each mass eigenstate separately is not a
discrete chiral invariant.   Such structures, if present in the low energy 
effective action, would violate the assertion of the Coleman-Mandula [20] 
theorem that the only symmetries of $S$-matrix are the direct product of the 
Poincar\'e group and an internal symmetry group.  So we are justified in 
excluding them {\it if} the Coleman-Mandula theorem extends to the case 
of discrete internal symmetry groups, such as the discrete chiral 
transformations of Eqs.~(13, 14).

An examination of the existing proofs of the Coleman-Mandula theorem shows 
that they make essential use of a continuity assumption,\footnote{*}{I am  
indebted to L. P. Horwitz for email correspondence on this point.} 
and so apply only 
to the case of a continuous internal symmetry group.  To see whether it 
is reasonable to postulate their extension to the discrete symmetry case, 
we have tried to construct a local Lagrangian counterexample.  Consider, for 
example a Lagrangian density containing the terms
$$\Psi_{ML}^{\dagger} \Phi_L^{2M+1}+\Psi_{NL}^{\dagger}\Phi_L^{2N+1}+...
~~~,\eqno(C2a)$$
with Lorentz indices contracted with each other or with partial derivatives 
to form a Lorentz scalar.  When the auxiliary field $\Phi_L$ undergoes 
a discrete chiral transformation 
$$\Phi_L \rightarrow \Phi_L \exp(i \alpha)~~~,\eqno(C2b)   $$ 
the invariance of the Lagrangian of Eq.~(C2a) requires that $\Psi_{ML}$  
transform as 
$$\Psi_{ML}\rightarrow \Psi_{ML} \exp[i2\pi(2M+1)\alpha]~~~,\eqno(C2c)$$       
     
and similarly for $\Psi_{NL}$.  Thus the existence of a Lagrangian of the 
form of Eq.~(C2a) would give a local Lagrangian 
example of the behavior leading 
to the off-diagonal mass term of Eq.~(C1a).  However, since $\Phi_L$ has 
only two components and is a local fermion field, any power of this field 
higher than the second vanishes, and so Eq.~(C2a) contains the two 
indicated terms only when both $M$ and $N$ are zero, which does not lead to 
any of the off-diagonal couplings discussed above.  To avoid having high
powers of the auxiliary field $\Phi_L$, one could try to introduce multiple 
auxiliary fields, and to construct a Lagrangian multilinear in these 
auxiliary fields so that the auxiliary fields are all forced to have 
the same chiral transformation of Eq.~(C2b) and $\Psi_{ML}$ is 
forced to have the chiral transformation of Eq.~(C2c).  The problem here 
is that by introducing multiple auxiliary fields one introduces the 
possibility of additional symmetries of the Lagrangian, that typically  
act to spoil the counterexample.  For example, consider the Lagrangian 
(with all fields now understood to be left handed, so the subscript $L$ is 
suppressed)
$$A\Phi_a^{\dagger}\Phi_c+B\Phi_b^{\dagger}\Phi_c +C\Phi_a^{\dagger}\Phi_b
+D\Psi_1^{\dagger}\Phi_a\Phi_b\Phi_c ~+~{\rm adjoint}~~~,\eqno(C3a)$$
that would seem to have all properties needed to force $\Psi_1$ to obey 
Eq.~(C2c) with $M=1$.  But redefining $\Phi_a \rightarrow A^{*-1} \Phi_a$ and 
$\Phi_b \rightarrow B^{*-1} \Phi_b$, Eq.~(C3a) takes the form 
$$\Phi_a^{\dagger}\Phi_c+\Phi_b^{\dagger}\Phi_c+C^{\prime} \Phi_a^{\dagger} 
\Phi_b +D^{\prime}\Psi_1^{\dagger}\Phi_a\Phi_b\Phi_c~+~{\rm adjoint} ~~~. 
\eqno(C3b)$$
The first three terms of Eq.~(C3b)  can be rewritten 
(including adjoints) as 
$$(\Phi_a+\Phi_b)^{\dagger}\Phi_c~+~{\rm adjoint} + 
{C^{\prime} \over 2} [(\Phi_a+\Phi_b)^{\dagger} (\Phi_a+\Phi_b) 
-(\Phi_a-\Phi_b)^{\dagger}(\Phi_a-\Phi_b)]~~~,\eqno(C3c)$$
showing that the chiral phase of $\Phi_a -\Phi_b$ is in fact not restricted 
by these terms. Thus the Lagrangian of Eq.~(C3a)  has a two parameter, 
rather than a one parameter,  
chiral symmetry group and so does not provide a suitable counterexample.   
The above arguments are not 
systematic enough to constitute a proof, but the fact that it does not seem 
easy to make a local Lagrangian counterexample suggests that the 
Coleman-Mandula theorem may be extendable to discrete chiral symmetries.  
This is an interesting question for further study.

\vfill\eject
\centerline{\bf References}
\bigskip
\noindent
\item{[1]} G. 't Hooft, in Recent developments in gauge theories, 
G. 't Hooft et al., eds. (Plenum, New York, 1980).  For an extensive 
bibliography, see ref. 4 of L. Alvarez-Gaum\'e and E. Witten, 
Nucl. Phys. B234 (1983) 269.  \hfill\break
\bigskip 
\noindent
\item{[2]}  C. H. Albright, Phys. Rev. D24 (1981) 1969; A. N. Schellekens,  
K. Kang, and I.-G. Koh, Phys. Rev. D26 (1982) 658. \hfill\break 
\bigskip
\noindent
\item{[3]}  N. Seiberg, Phys. Rev. 49 (6857) 1994; hep-th/9402044. 
\hfill\break
\bigskip
\noindent
\item{[4]}
N. Seiberg, ``The Power of Holomorphy - Exact Results in 
4D SUSY Field Theories'', hep-th/9408013, Secs. 3 and 5.  See also 
K. Intriligator and N. Seiberg, ``Lectures on Supersymmetric 
Gauge Theories and Electric-Magnetic Duality'',hep-th/9509066, Secs. 4.3 and 
5.4; to appear in the Proceedings of the Trieste '95 spring school, 
TASI '95, Trieste '95 summer school, and Cargese '95 summer school.  
\bigskip
\noindent
\item{[5]}  H. Harari and N. Seiberg, Phys. Lett. 98B (1982) 269; 
Nucl. Phys. B204 (1982) 141. \hfill\break 
\bigskip
\noindent
\item{[6]}  H. Harari, Phys. Lett. 86B (1979) 83; M. A. Shupe, Phys. Lett. 
86B (1979) 87.  \hfill\break
\bigskip
\noindent
\item{[7]}  S. L. Adler, Quaternionic quantum mechanics and quantum fields 
(Oxford University Press, New York, 1995).\hfill\break
\bigskip
\noindent
\item{[8]}  R. N. Mohapatra, Unification and Supersymmetry, Second Edition 
(Springer, New York, 1992), Chapt. 6. \hfill\break
\bigskip
\noindent
\item{[9]}  R. Slansky, Phys. Reports 79 (1981) 1.  \hfill\break
\bigskip
\noindent
\item{[10]}  D. Z. Freedman, Review of supersymmetry and supergravity, in 
Proceedings of the XIXth International Conference on High Energy Physics, 
Tokyo, 1978 (Physical Society of Japan, Tokyo, 1979), p. 535.  \hfill\break
\bigskip
\noindent
\item{[11}  J. Pati and A. Salam, Phys. Rev. D10 (1974) 75; 
J. Pati, Twenty years later: why I still believe in $SU(4)$, talk at 
Salamfest, Trieste, March, 1993.  \hfill\break
\bigskip
\noindent
\item{[12]}  J. Strathdee, Int. Journ. Mod. Phys. A2 (1987) 273.\hfill\break  
\bigskip
\noindent
\item{[13]}  D. Chowdhury, Spin glasses and other frustrated systems,  
(Princeton University Press, Princeton, 1986), Sec.~1.3; M. Mezard, 
G. Parisi, 
and M. A. Virasoro, Spin glass theory and beyond (World Scientific, 
Singapore, 
1987), Chapt.~0.\hfill\break
\bigskip
\noindent
\item{[14]}  See, e.g., p.146 of the July 1996 Particle Physics Booklet 
abstracted from R. M. Barnett et. al., Phys. Rev. D54 (1996) 1.\hfill \break
\bigskip
\noindent
\item{[15]}  M. E. Peskin, Chiral symmetry and chiral symmetry breaking, 
in Recent advances in field theory and statistical mechanics (Les Houches,  
1982), J.-B. Zuber and R. Stora, eds. (North-Holland, Amsterdam, 1984).  
See also J. Kiskis, Phys. Rev. D15 (1977) 2329; L. S. Brown, R. D. Carlitz, 
and C. Lee, Phys. Rev. D16 (1977) 417; S. Coleman, Aspects of symmetry 
(Cambridge University Press, Cambridge, 1985), Chapt. 7. \hfill\break
\bigskip
\noindent
\item{[16]}  I-H. Lee and R. Shrock, Phys. Lett. B201 (1988) 497; 
S. Aoki, I-H. Lee,  and R. Shrock, Phys. Lett.  B207 (1988) 471.\hfill\break
\bigskip
\noindent
\item{[17]}  J. M. Cornwall, Phys. Rev. D10 (1974) 500; S. Raby, 
S. Dimopoulos, and L. Susskind, Nucl. Phys. B169 (1980) 373.\hfill\break  
\bigskip
\noindent
\item{[18]}  S. Weinberg, Phys. Lett. 102B (1981) 401. \hfill\break
\bigskip
\noindent
\item{[19]}  H. Harari and N. Seiberg, Phys. Lett. 102B (1981) 263.  
\hfill\break
\bigskip
\noindent
\item{[20]}  S. Coleman and J. Mandula, Phys. Rev. 159 (1967) 1251; O. Pelc 
and L. P. Horwitz, J. Math. Phys. 38 (1997) 1.\hfill\break  
\bigskip
\noindent
\item{[21]}  A. Zee, Phys. Lett. 95B (1980) 290; R. N. Mohapatra, Ref. 6, 
Chapt. 8.  \hfill\break
\bigskip
\noindent
\item{[22]}  W. G. McKay and J. Patera, Tables of dimensions, indices, and 
branching rules for representations of simple Lie algebras (Marcel 
Dekker, New York and Basel, 1981), Table 2, p. 86 for representations 
of $A3$ or $SU(4)$ and p. 138 for the branching rules of representations of 
$A5$ or $SU(6)$.  
\bigskip
\noindent
\item{[23]}  B. Block and H. Feshbach, Ann. Phys. (NY) 23 (1963) 47;    
H. Feshbach, A. K. Kerman, and R. H. Lemmer, Ann. Phys. (NY) 41 (1967) 230. 
\hfill\break
\bigskip
\noindent
\item{[24]}  D. Chang, R. N. Mohapatra, and M. K. Parida, Phys. Rev. Lett. 
50 (1984) 1072; Phys. Rev. D30 (1984) 1052.  \hfill\break
\bigskip
\noindent
\item{[25]}  S. Coleman, J. Wess, and B. Zumino, Phys. Rev. 177 (1969) 2239; 
C. G. Callan, Jr., S. Coleman, J. Wess, and B. Zumino, Phys. Rev. 177 (1969) 
2247. \hfill\break
\bigskip
\noindent
\item{[26]}  S. Elitzur, Phys. Rev. D12 (1975) 3978. \hfill\break
\bigskip
\noindent
\item{[27]}  For a review, see T. Applequist, Dynamical electoweak 
symmetry breaking, YCTP-P23-91, 1991. \hfill\break
\bigskip
\noindent
\item{[28]}  H. Harari, R. N. Mohapatra, and N. Seiberg, Nucl. Phys. 
B201 (1982) 174.  \hfill\break
\bigskip
\noindent
\item{[29]} E. Bergshoeff, M. De Roo, and B. De Witt, Nucl. Phys. B182  
(1981) 173; E. Fradkin and A. A. Tseytlin, Phys. Lett. 134B (1984) 187.  
\hfill\break
\bigskip 
\noindent
\item{[30]} P. Langacker and J. Erler, Sec. 14 in R. M. Barnett et. al., 
Review of particle physics, Phys. Rev. D54 (1966) 1, pp. 103-106.  
\hfill\break
\bigskip
\noindent
\item{[31]}  M. J. Dugan and L. Randall, Phys. Lett. B264 (154) 1991.  
\hfill\break
\bigskip
\noindent
\item{[32]} H1 Collaboration, DESY 97-025, and ZEUS Collaboration, DESY 
97-24, submitted to Zeitschrift f\"ur Physik C.  \hfill\break
\bigskip
\noindent
\item{[33]} R. Slansky, T. Goldman, and G. L. Shaw, Phys. Rev. Lett. 
47 (1981) 887.
\hfill\break
\bigskip
\noindent
\item{[34]} K. Akama, K. Katsuura, and H. Terazawa, hep-th/9704327.   
\hfill\break
\bigskip
\noindent
\item{[35]}  D. Fairlie, J. Nuyts, and A. Taormina, Phys. Rev. D27 (1983) 
264.  
\hfill\break
\hfill\break
\vfill
\eject
\bigskip
\centerline{\bf Figure Caption}
\medskip
\item{Fig. 1}  Matching of the two branches of the running coupling 
across the nonperturbative regime around $\Lambda_H^2$.  The match is 
parameterized by $g_*^2$, the value of the coupling at the intersection  
of the extrapolated $g^2_{SU(3)}$ of branch 2 with $g^2_{SU(4)}$ of 
branch 1.  The magnitude of $g_*^2$ is determined by the physics of 
bound state formation and gauge charge screening in the nonperturbative 
regime.  
\vfill
\eject
\bigskip
\bye